\documentclass[preprint,authoryear,12pt]{elsarticle}
\pdfoutput=1
\usepackage{amsmath,amssymb}
\usepackage{verbatim}
\usepackage{hyperref}
\usepackage{cleveref}
\usepackage{graphicx}
\usepackage{numprint}
\usepackage{natbib}
\usepackage{tikz}
\usetikzlibrary{shapes,arrows,decorations.pathmorphing,backgrounds,positioning,fit,calc,shadows}
\usepackage{rotating}
\usepackage{subfigure}
\graphicspath{{./}{pdffigs/}}
\journal{Computational Statistics and Data Analysis}

\newcommand{\margnote}[1]{ }

\begin{document}

\begin{frontmatter}

\title{Clustering in networks with the collapsed Stochastic Block Model}
\date{September 22 2012}

\author{
	Aaron F. McDaid\footnote{Correspondence to: CASL UCD, 8 Belfield Office Park, Clonskeagh, Dublin 4, Ireland. \emph{Email:} aaronmcdaid@gmail.com. \emph{Tel:} +35385775686 }
, Thomas Brendan Murphy, Nial~Friel and Neil J. Hurley \\
       Clique Research Cluster \\
       University College Dublin\\
}

\address{}

\begin{abstract}
	An efficient MCMC algorithm is presented
	to cluster the nodes of a network such that nodes with
	similar role in the network are clustered together.
	This is known as \emph{block-modelling} or \emph{block-clustering}.
	The model is the stochastic blockmodel (SBM)
	with block parameters integrated out.
	The resulting marginal distribution defines a posterior over the number of clusters and cluster memberships.
	Sampling from this posterior is simpler than from the original SBM as transdimensional MCMC can be avoided.
	The algorithm is based on the \emph{allocation sampler}. 
	It requires a prior to be placed on the number of clusters, thereby allowing the
	number of clusters to be directly estimated by the algorithm, rather than being given as an input parameter.
	Synthetic and real data are used to test the speed and accuracy of
	the model and algorithm, including the ability to estimate the number of clusters.
	The algorithm can scale to networks with up to ten thousand nodes and tens of millions of edges.
\end{abstract}


\end{frontmatter}

\newcommand{\pr}{\mathrm{P}}
\newcommand{\B}{\mathrm{B}}
\newcommand{\z}{ \ensuremath{ \mathbf{z} } }
\newcommand{\x}{ \ensuremath{ \mathbf{x} } }
\newcommand{\apriori}{\textit{a priori}}
\newcommand{\Z}{ \ensuremath{ \mathbb{Z} } }
\newcommand{\betabinom}[4]{ \frac{ \B(#1 + #3,#2 - #1 + #4) }{ \B(#3,#4)} }
\newcommand{\comments}[1]{}
\newcommand{\argmax}[1]{\ensuremath{\underset{ #1  }{\operatorname{arg\,max}} \, }}

\npthousandsep{,}\npthousandthpartsep{\npdecimalsign{.}}


\tikzset{vertex/.style={circle,draw=gray!70,fill=gray!30} }
\tikzset{invertex/.style={vertex,black} }
\tikzset{edge/.style={shorten <= 1pt, shorten >= 1pt,black,ultra thick} }
\tikzset{drawn/.style={edge,color=blue!50,arrows=->} }

\section{Introduction}

This paper
is concerned with \emph{block-modelling} -- an approach to clustering the nodes in a network, based on the pattern of inter-connections between them.
\margnote{Rewritten these first two paragraphs, so we don't claim to have invented any extensions.}
The starting point for the method presented here is the \emph{stochastic block model} (SBM) \cite{Nowicki-01}.
The goal is to improve the speed and scalability, without compromising on accuracy.
We use conjugate priors and integration in order to focus on the marginal distribution of interest, 
this marginalization is also referred to as the `collapsing' of the nuisance parameters \citep{LiuCollapsedGibbs, WyseFriel}.
This allows us to implement an efficient algorithm based on the \emph{allocation sampler} of \cite{NobileAllocationSampler}.
We incorporate existing extensions, such as the weighted-edge model of \cite{MariadassouWeightedSBM}, and show
how this extension can be incorporated within our collapsing and within our algorithm.
As required by the \emph{allocation sampler}, we place a prior on the number of clusters, allowing the number of clusters to be directly estimated.
Together, these techniques allow us to avoid the more complex forms of transdimensional MCMC and they also allow us
to avoid the need for post-hoc model selection via criteria such as the ICL.

We show that our method can accurately and efficiently estimate the number of clusters -- an improvement over many existing methods.
Our
algorithm, and the data we have used in \cref{SEClargernetworks} and our survey data used in \cref{SECsurvey}, are available at \url{http://sites.google.com/site/aaronmcdaid/sbm}.

The concept of clustering is broad and originated outside of network analysis, where the input data is in the form of real-valued vectors  describing the location of the data points in a Euclidean space.  Network clustering takes a set of connected nodes as input  and finds a partition of the nodes based on the network structure.  This finds application in many different contexts.  For instance, in bio-informatics, networks of protein-protein interactions are analysed and clustering is applied to find functional groups of proteins.  Interest in social network analysis has grown greatly in recent years, with the availability of many networks, such as Facebook datasets, of human interactions. Clustering of such social networks has been applied in order to find social communities. In the following, we will distinguish the community-finding problem from the more general setting of block-modelling. 

In network analysis, the  input data may be described mathematically as a graph, which is a set of nodes
(where each node represents an entity, say, a person) and a set of edges linking pairs of nodes together. 
An edge might represent a friendship on Facebook or a phone call on a mobile phone network. In  \cref{SECsurvey}, we apply our method to the network of interactions between participants at a summer school.

Given a network, the goal in block-modelling is to cluster the nodes such that pairs of nodes are
clustered together if their connectivity pattern to the clusters in the rest of the network is similar. A
cluster might, for example, consist of a set of nodes which do not tend to have connections among themselves at all.
Given two nodes in this cluster (node  $i$ and node $j$), the neighbours of $i$ tend to be in the same clusters as are the neighbours of $j$.
Community-finding has focussed on finding clusters of high internal edge density,
where an edge between two nodes will tend to pull the two nodes into the same cluster, and a non-edge will tend to push them into separate clusters.
This contrasts with block-modelling, which allows clusters to have \emph{low} internal edge density.
Block-modelling is able to find such community structure, but it is a more general method that
is also able to find other types of structure.

A variety of other, non-probabilistic, approaches have been used to tackle the broad problem
\margnote{More refs, for Rev\#1.  But no discussion, to satisfy Rev\#2. Thoughts?}
of block-modelling \citep{EverettColoration,ChanGBM}.
Outside of block-modelling, there are other solutions for community-finding in networks \citep{NewmanGirvan,girvan-2002,NewmanFastishMod}.
Many probabilistic clustering models have also been applied \citep{HandcockRafteryTantrum,HoffLatentSpace,Airoldi2008MMSB}.

There is a huge variety of methods, and we will not attempt to summarize them further;
for the rest of this paper, we will focus on the SBM and on algorithms for the SBM.
For more details, in particular about community finding, see the excellent review article of \cite{fortunato-2010}.

The remainder of the paper is structured as follows. In \cref{SECsbm}, we define the SBM and define the notation used in the paper.
We then define, in \cref{SECours}, the conjugate priors and integration that we use in order to access the relevant marginal distribution.
 \Cref{SECrelatedProb} discusses other closely-related models and algorithms and in particular gives consideration to the issue of how
to select the number of clusters (model selection), comparing
the approach we have used to other approaches and noting connections among the methods.

\Cref{SECestimation} describes the algorithm we use;
\margnote{Rev\#2 is \emph{very} unhappy with the term `collapsing' - see his comments (in red) on page 40 below.}
without collapsing, it would have been necessary to use full
Reversible Jump MCMC (\cite{GreenRJMCMC}) to search a sample space of varying dimension and this could be much slower.

In \cref{SECevaluation}, we evaluate our method on synthetic networks, showing how the number of clusters can be estimated accurately and the nodes assigned to
their correct cluster with high probability.
We also test the scalability and efficiency of the algorithm by considering synthetic datasets with ten thousand nodes and ten million edges.

In \cref{SECsurvey}, we evaluate our method on a dataset of interactions, gathered by a survey, of participants at
a doctoral summer school attended by one of the authors of this paper.
The method is able to detect interesting structures, demonstrating the differences between \emph{block-modelling} and \emph{community-finding}. 
\Cref{SECconclusion} draws some conclusions.

\section{Stochastic Block Model(SBM)}
\label{SECsbm}
As formulated in \cite{Nowicki-01}, a network describes a relational structure on a set of nodes. Each edge in the network describes a relationship between the two nodes it links.
A general case of a finite alphabet of states relating a pair of nodes is considered but in the simplest case, discussed by the same authors in \cite{SnijdersSBM1997}, relationships are binary -- an edge joining a pair of nodes either exists or not. 
The network can be undirected, corresponding to symmetric relationships between the nodes, or may be \emph{directed}, where a relationship from node $i$ to node $j$ does not necessarily imply the same relationship exists from node $j$ to node $i$. Finally, a \emph{self-loop} -- a relationship from node to itself -- may or may not be allowed.

Throughout the paper, we use $\mathrm{P}(\cdot)$ to refer to probability mass (i.e. of discrete quantities)
and $\mathrm{p}(\cdot)$ to refer to probability density (i.e. of continuous quantities).
$N$ is the number of nodes in the network and $K$ is the number of clusters.
In the algorithm proposed in \cite{Nowicki-01}, these are given input values, although 
in our approach, we 
treat $K$ as a random variable with a given prior distribution.
Given $N$ and $K$, the SBM  describes a random
process for assigning the nodes to clusters and then generating a network. 
Specifically, the cluster memberships are represented by a random vector $z$ of length $N$  such that  $z_i \in \{1, \dots, K\}$ records the cluster containing node $i$.
$z_i$ follows a multinomial distribution,
\[ z_i \overset{iid}\sim \text{Multinomial}( 1; \theta_1, \dots, \theta_K) \, , \]
\noindent such that $\theta_i$ is the probability of a node being assigned to cluster $i$ ($1 = \sum_{k=1}^K \theta_k$). 
The vector $\theta$ is itself a random variable drawn from a Dirichlet prior with dimension $K$.
The parameter to the Dirichlet is a vector $(\alpha_1,\dots,\alpha_K)$ of length $K$. We follow \cite{Nowicki-01} by fixing the components of this vector to a single value $\alpha$, and by default $\alpha = 1$,
\[ \theta \sim \text{Dirichlet}(\alpha_1 = \alpha, \alpha_2 = \alpha, \dots, \alpha_K = \alpha ) \,.\]

This describes fully how the $N$ nodes are assigned to the $K$ clusters.
Next we describe how, given this clustering $z$, the edges are added
between the nodes.

A network can be represented as an $N \times N$ adjacency matrix,
$x$, such that $x_{ij}$ represents the relation between node $i$ and node $j$ (taking values 1 or 0 in the binary case).
Denote by $x_{(kl)}$ the submatrix corresponding to the \emph{block} of connections between nodes in cluster $k$ and nodes in cluster $l$.
If the network is undirected, there are $\frac1 2 K (K+1)$ blocks, corresponding to each pair of clusters;
and if the network is directed, there are $K^2$ clusters, corresponding to each \emph{ordered} pair of clusters.

It is generally simpler to discuss the directed model; unless otherwise stated, the formulae presented here apply only to the directed case.
The definitions and derivations can easily be applied to the undirected case, provided that care is taken only
to consider each pair of nodes exactly once.

If self-loops are not allowed, then the diagonal entries of $x$, $x_{ii}$, are excluded from the model. It is assumed that, given $K$ and $z$, connections are formed independently within a block so that
\[
	P(x | z, K, \pi) = \prod_{k.l} P(x_{(kl)} | z, K, \pi_{kl}) \, ,
\]
where
\[
          P(x_{(kl)} | z, K, \pi_{kl}) = \prod_{\{i | z_i = k\}} \prod_{\{j | z_j = l\}} P(x_{ij} | z, K, \pi_{kl})\,,
\]
and the matrix $\pi = \{\pi_{kl}\}$ describes the cluster-cluster interactions.
$\pi$ is a $K\times K$ matrix, but for undirected networks only the diagonal and upper triangle are relevant.
Specifically, for binary networks, $\pi_{kl}$ represents the edge density within the block, and edges follow the Bernoulli distribution,
\[ x_{ij} | z, K, \pi \sim \mathrm{Bernoulli}(\pi_{z_i z_j}) \,. \]
Each of the $\pi_{kl}$  is drawn  from the conjugate  $\text{Beta}(\beta_1, \beta_2)$ prior,
\[
\pi_{kl} \overset{iid}\sim \text{Beta}(\beta_1, \beta_2)\,.
\]
Again we follow \cite{Nowicki-01} and choose $\beta_1=\beta_2=1$, giving a Uniform prior.

This completes the description of the Bayesian presentation of the SBM.
A different approach is taken in other work, such as that of \cite{Daudin-08},
\margnote{
	Both reviewers want to discuss freq-vs-Bayes.  I'm trying to satisfy them while also focussing on the practical differences (which are quite small)
	instead of the philosophy.
	The `frequentist' SBM methods don't use unbiased estimators or any of the machinery of a full-blown frequentist.
	In practice, the methods give the same point estimates, \emph{if} we were to use a Bayesian point estimate.
}
where, using essentially the same model, the goal is to take a point estimate of the parameters, $(\pi,\theta)$, for a given number of clusters $K$.
Specifically, the aim is to find the MLE; the value of $(\hat\pi,\hat\theta)$ which maximizes $\mathrm{P}(x | \pi, \theta, K)$.
This is described as the frequentist approach, in contrast to the fully Bayesian approach
where a distribution of parameter values is allowed
instead of a point estimate.
We will return to this issue in a little more detail in \cref{SECrelatedProb} in order
to discuss the practical differences from an algorithmic point of view.

\subsection{Data model variations}
The model is  naturally extended in \cite{Nowicki-01}  to allow for a finite alphabet of two or more relational states, where instead of using a Bernoulli with a Beta prior for $x$ and $\pi$,
we can use a Multinomial and a Dirichlet  to model this alphabet.
The Bernoulli-and-Beta-prior model is just a special case of the
Multinomial-and-Dirichlet-prior model.
Alternatively, we can allow an infinite support and extend the model to allow for non-negative integer weights on the edges,
by placing a Poisson distribution on $P(x|\pi , z)$, as seen in \cite{MariadassouWeightedSBM}.
Now $\pi_{kl}$ represents the edge rate and is drawn from a Gamma prior,
\[
\begin{split}
	x_{ij} | z, K, \pi & \sim \text{Poisson}(\pi_{z_i z_j}) \\
\pi_{kl} 
& \sim \mbox{Gamma}(s, \phi) \, .
\end{split}
\]
We do not suggest any default for the hyperparameters $s$ and $\phi$.  
A further extension to real-valued weights is also possible, by using a Gaussian
for $p(x|\pi,z)$ and suitable prior on $\pi$, following \cite{WyseFriel}.
These variations, and others, are described in \cite{MariadassouWeightedSBM},
but they do not discuss conjugate priors.
\margnote{ Rewritten subsection, so as not to claim that we've discovered anything new.}

The integration approach and algorithm described later in this paper can be applied to many variants of edge model, however we focus in the remainder of the paper on the Bernoulli and Poisson models that are supported in our software.

In summary, given $N$ and $K$ the random process generates $\theta$, $z$, $\pi$ and ultimately  the network $x$.
The two main variables of interest are the clustering $z$ and the network $x$. In a typical application, we have observed a network $x$ and perhaps we have an estimate of $K$, and our goal is to estimate $z$.




\section{Collapsing the SBM}
\label{SECours}

In this section, we show how \emph{collapsing} can be used to give a more convenient and efficient expression for the model.
This refers to the integration of nuisance parameters from the model, see \cite{WyseFriel} for an application to a different,
but related, bipartite model.
The SBM has been partially collapsed by \cite{KempTRcollapsedSBM}, but we will consider the
full collapsing of both $\pi$ and $\theta$.
As our primary interest is in the clustering $z$ and the number of clusters $K$,
we integrate out $\pi$ and $\theta$,
yielding an explicit expression for the marginal $\mathrm{P}(x, z, K)$.
We emphasize that integration  does not change the model, it merely yields
\margnote{Collapsing is \emph{not} an extension, as pointed out by Rev\#2.
	Perhaps there is a word, other than `extension', that we want to use? I'm happy as is.
}
a more convenient representation of the relevant parts of the posterior.
This integration is made possible by the choice of conjugate priors for  $\pi$ and $\theta$.
We treat $K$ as a random variable and place a Poisson prior on $K$ with rate $\lambda=1$, conditioning on $K>0$,
\begin{equation}
	K \sim \mbox{Poisson}(1) \, | \, K>0 \, ,
	\label{EQpriorK}
\end{equation}
which gives us
\[
\mathrm{P}(K) = \frac{ \frac{\lambda^K}{K!} e^{-\lambda} }{ 1-\mathrm{P}(K=0) }
              =        \frac{1}{K!(e-1)}                    \,.
      \]
We are only interested in these expressions as functions of $K$ and $z$ up to proportionality,
as this will be sufficient for our Markov Chain over $(K,z|x)$,
and hence we can simply use $\mathrm{P}(K) \propto \frac1{K!}$.

The Poisson prior is used in the \emph{allocation sampler}, the algorithm upon which our method is based \citep{NobileAllocationSampler}.
This allows the estimation of the
number of clusters as an output of the model rather than requiring a user to specify $K$  as an input  or to
to use a more complex form of model selection.
Thus, we have a fully Bayesian approach where, other than $N$, which is taken as given, every other quantity is a random variable with specified priors where necessary,
\begin{equation}
	\begin{split}
\mathrm{p}(x, \pi, z, \theta, K) =
\mathrm{P}(K)
& \times \mathrm{p}(z,  \theta | K)  \\
& \times \mathrm{p}(x , \pi | z) \, .
	\end{split}
\label{EQuncollapsed}
\end{equation}  

With \cref{EQuncollapsed} we could create an algorithm which, given a network $x$, would
allow us to sample the posterior $\pi, z, \theta, K | x$.
However, we are only interested in estimates of $z,K | x$.
We now show how to collapse $\pi$ and $\theta$.

Define $\mathbb{R}_+$ to be the set of non-negative real numbers, 
and write the set of real numbers between 0 and 1 as $[0,1] $. 
Define $\Theta$ the \emph{unit simplex} i.e.  the subset of $\mathbb{R}_+^K$ where $1=\sum_{k=1}^K \theta_k$.
Define $\Pi$ to be the domain of $\pi$. For the Poisson model this is $\mathbb{R}_+^B$ while
for the Bernoulli model this is $[0,1]^B$, where $B$ is the number of blocks.

We can access the same posterior for $z$ and $K$ by \emph{collapsing} two of the factors in \cref{EQuncollapsed},
\begin{equation}
	\begin{split}
\mathrm{P}(x, z, K) =
\mathrm{P}(K)
\times \int_\Theta \mathrm{p}(z,  \theta | K) \; \mathrm{d}\theta   \\
\times \int_\Pi    \mathrm{p}(x , \pi | z)    \; \mathrm{d}\pi   \, ,
	\end{split}
\label{EQcollapsed}
\end{equation}
or, equivalently, using the block-by-block independence $x_{(kl)} | z,K$,
\begin{equation}
	\begin{split}
\mathrm{P}(x, z, K) =
\mathrm{P}(K)
\times \int_\Theta \mathrm{p}(z,  \theta | K) \; \mathrm{d}\theta   \\
\times \prod_{k,l} \int_{\Pi_{kl}}    \mathrm{p}(x_{(kl)} , \pi_{kl} | z)    \; \mathrm{d}\pi_{kl} \, .
	\end{split}
\label{EQcollapsed2}
\end{equation}

This allows the creation of an algorithm which searches only over $K$ and $z$.
The algorithm never needs to concern itself with $\theta$ or $\pi$.

Collapsing
greatly simplifies the sample space over which the MCMC algorithm has to search.
Without collapsing, the dimensionality of the sample space would change if our estimate of $K$ changed;
this would require a Reversible-Jump Markov Chain Monte Carlo (RJMCMC) algorithm (see \cite{GreenRJMCMC}).
\margnote{Per Rev\#2, I've deleted the argument about how collapsing will improve mixing.  This is quite frustrating.}
Finally, if estimates for the
full posterior, including $\pi$ and $\theta$, are required, it should
be noted that it is very easy to sample $\pi,\theta | x,z,K$,
meaning that nothing is lost by the use of collapsing.
Many of the other models described in \cref{SECrelatedProb} are collapsible,
and this may be an avenue for future research.

The integration of \cref{EQcollapsed2}  allows an expression for the full posterior distribution to be obtained.
Details of the derivation of this expression  are given in Appendix A. 
Let $n_k$ be the number of nodes in cluster $k$. $n_k$ is a function of $z$.
For the Bernoulli model, let $y_{kl}$ be the number of edges that exist in block $kl$,  i.e. the
block between clusters $k$ and $l$.  For the Poisson model, $y_{kl}$ is
the total edge weight.  $y$ is a function of $x$ and $z$.
Let $p_{kl}$ be the maximum number of edges that can be formed between clusters $k$ and $l$.
For off-diagonal blocks, $p_{kl} = n_k n_l$. For diagonal blocks, $p_{kk}$ depends on the form of the network as follows,

\begin{equation}
	p_{kk} = \left\{ \begin{array}{ll}
                        \frac12 n_k (n_k-1) & \mbox{undirected, no self-loops} \\
                        \frac12 n_k (n_k+1) & \mbox{undirected, self-loops} \\
                        n_k (n_k-1) & \mbox{directed, no self-loops}\\
                         n_k^2 & \mbox{directed, self-loops}
                       \end{array}
		       \right.
		       \, .
\label{EQpkkCountPairs}
\end{equation}
The full posterior may be written as

\begin{equation}
	\begin{split}
\mathrm{P}(x, z, K)
\propto {}
&
\frac1{K!}  \\
& \times \frac
{\Gamma(\alpha K) \prod_{k=1}^K \Gamma(n_k + \alpha)}
{\Gamma(\alpha)^K \Gamma(N + \alpha K)}
\\
& \times \prod f(x_{(kl)} | z) \,,
	\end{split}
\label{EQfinal}
\end{equation}
where the final product is understood to take place over all blocks. The  form of the function $f(x_{(kl)} | z)$
depends on the edge model.  If Bernoulli, then
\begin{equation}
f(x_{(kl)} | z) = \frac{ \text{B}(\beta_1 + y_{kl}, p_{kl} - y_{kl} + \beta_2) }{ \text{B}(\beta_1, \beta_2) }\,,
\label{EQfBernoulli}
\end{equation}
where $B(.,.)$ is the Beta function.
If Poisson, then
\begin{equation}
	f(x_{(kl)} | z) = \frac{ \Gamma(s+y_{kl}) \left( \frac1{p_{kl} + \frac1\phi} \right)^{s+y_{kl}} }{ \Gamma(s) \phi^s }\,.
\label{EQfPoisson}
\end{equation}


\section{Related estimation procedures for the SBM}
\label{SECrelatedProb}
\margnote{This section has been shortened a lot, as per Rev\#2. It now focusses strictly on the SBM, and very-closely-related models.
I start with the models that are most different from ours, and end with those that are most similar.
I no longer say very much about their algorithms; I limit myself to their overall goal and strategy.
}

Before defining our algorithm, we look at
related work, particularly other methods that are based
on the SBM.
We will focus on models which are identical, or very similar to, the SBM.
Therefore, we will not discuss other models which are loosely related, such as
that of \cite{newman-2007}, or the ``degree-corrected'' SBM of \cite{KarrerDegreeCorrected}.

All methods discussed here are aimed at estimating $z$, but they differ in the approach they take to the parameters $\pi$ and $\theta$ and
in whether they allow the number of clusters, $K$, to be estimated.
We also  discuss the issue of model selection, i.e. how the various methods
estimate the number of clusters.
This question was avoided in the original paper of \cite{Nowicki-01}, where the number of clusters is fixed to $K=2$ in the evaluation.

The method of \cite{Daudin-08} takes a network, $x$, and number of clusters $K$,
and applies a variational algorithm.
Point estimates are used for $\pi$ and $\theta$, but the clustering $z$ is represented as a distribution
of possible cluster assignments for each node. This makes the method analogous to the EM
algorithm for the MLE -- finding the pair $(\pi,\theta)$ which maximizes $\mathrm{P} (x | \pi, \theta, K)$.

The model used by \cite{ZanghiOnline} is a subset of the model of \cite{Daudin-08}.
The cluster-cluster density matrix, $\pi$,
is simplified such that it is represented by two parameters $\lambda$ and $\epsilon$, such that the on-diagonal blocks $\pi_{kk} = \lambda$
and the off-diagonal blocks  $\pi_{kl} = \epsilon$ (for $k \neq l$).
A Classification EM (CEM) algorithm  to maximize
$ \underset{z,\pi,\theta}{\operatorname{arg max}} \; \mathrm{P} (x,z | \pi, \theta, K)$ is briefly described in \cite{ZanghiOnline} but not implemented.
Instead, they implement an \emph{online} algorithm.
One node of the network is considered at a time and is
assigned to the cluster which maximizes $\mathrm{P} (x,z | \pi, \theta, K)$, updating estimates of $\pi$ and $\theta$ with each addition.
Implicitly, their goal is to use point estimates both for the parameters \emph{and} for the clustering,
to find $(\hat{z},\hat\pi,\hat\theta)$ that would maximize $\mathrm{P}(x,z | \pi,\theta, K)$;
as such, it is loosely related to the profile likelihood \citep{BickelNonparametricView}.

The methods just discussed are based, directly or indirectly, on the frequentist approach of finding the maximum likelihood estimate of the parameters, $(\pi,\theta)$, i.e. the values
 $\hat\pi,\hat\theta$ that would maximize the likelihood of the observed network,
\[ \mathrm{P}(x | \pi,\theta, K) = \sum_{z} \mathrm{P}(x, z | \pi,\theta, K) \, . \]
The estimate of $z$ that is used in this frequentist approach is the conditional distribution of $z$ based
on this point estimate of the parameters and on the observed network,
$ z | x,\hat\pi, \hat\theta, K$.
In practice though, it is not tractable to calculate or maximize this likelihood exactly, and hence
a variety of different approximations and heuristics have been used.

In a Bayesian method, such as ours, a distribution of estimates for $(\pi,\theta)$ is used instead of a point estimate.
The goal is to directly sample from $z|x,K$.
Another example of this Bayesian approach is the variational algorithm used in \cite{HofmanBayesNetworkModularity},
which is based on the simpler $\lambda$ and $\epsilon$ parameterization of the $\pi$ matrix used in  \cite{ZanghiOnline}.

The modelling choices of \cite{LatoucheILvbStatMod}, where a new model selection criterion called $ILvB$ is introduced, are essentially identical to
the standard SBM;
each element of $\pi_{kl}$ is independent, and conjugate priors are specified.
A variety of other variational approximations are considered by \cite{GazalVariational},
where there is more focus on parameter estimation and less focus on model selection.

A further specialization of this model is possible, by employing the $\lambda, \epsilon$ parameterization, but where $\lambda > \epsilon$, which
explicity constrains the expected edge density within clusters 
to be larger than the expected edge density between clusters.
This can be considered to be \emph{community-finding} as opposed to \emph{block-modelling}.
The authors of this paper considered this in \cite{McDaidSCFcompstat}.

\subsection{Model selection}
\margnote{
	As per Rev\#2, this subsection is now much shorter.
It is now much more focussed on issues that are directly relevant to our algorithm.}
Later, in our experiments in \cref{SECevaluation}, we will demonstrate the ability of the allocation sampler to accurately estimate
the number of clusters.
In this subsection, we will briefly discuss some of the theoretical issues around the estimation of the number of clusters.

The methods that involve the MLE for the parameters involve the risk of overfitting;
for larger values of $K$, the parameter space of $\pi$ and $\theta$ becomes much larger
and therefore the estimates of $\mathrm{P}_{\theta = \theta_{mle}}(x | K)$ will become over-optimistic,
and will tend to overestimate $K$ \citep{schwarz1978}.
Therefore, an alternative formulation such as the ICL is needed;
see \cite{ZanghiOnline} and \cite{Daudin-08} for derivations of the ICL in the context of models based on the SBM.
Instead of using the MLE directly, those measures apply priors to the parameters and integrate over the priors,
as described in \cite{BiernackiICL}, such that the average likelihood is used instead of the maximum likelihood.

Typically, such integrations cannot be performed exactly and the ICL criterion consists of
approximations that are based on first finding an estimate to the MLE, and then adding correction terms to this MLE.
For the rest of this subsection, we will not consider those approximate methods
and will instead consider the exact solutions to the integrations.

The \emph{integrated~classification~likelihood}, which the ICL intends to approximate,
\[\mathrm{P}(x,z|K) = \int \int \mathrm{P}(x,z,\pi,\theta|K) \, \mathrm{d}\pi \, \mathrm{d}\pi \, , \]
can be solved exactly in some
models.  The SBM is one of those models, and the posterior mass that our algorithm
samples from is exactly equal to the integrated~classification~likelihood (if a uniform prior is used for $K$ instead of the default Poisson).
While it is easy to exactly calculate the integrated classification likelihood for a given $(z,K)$,
it would not be tractable to search across all possible $(z,K)$ to find
the state that maximizes the integrated classification likelihood, except for the smallest of networks.

The BIC is an attempt to approximate the \emph{integrated likelihood}
\[\mathrm{P}(x|K) = \sum_z \int \int \mathrm{P}(x,z,\pi,\theta|K) \, \mathrm{d}\pi \, \mathrm{d}\theta . \]
An exact solution to the BIC is not tractable for the SBM; the likelihood would require a summation over all possible clusterings $z$.

If we were to use a uniform prior for $K$,
then $\mathrm{P}(x|K) = \mathrm{P}(K|x)$ and
an irreducible ergodic Markov chain algorithm such as ours would visit each value of $K$
in proportion to the integrated likelihood for that value of $K$.
Of course, our algorithm only gives a \emph{sample} from the true posterior,
and there cannot be any guarantee that the distribution of the sample is
representative of the true distribution.

The purpose of these last few paragraphs is to demonstrate that there
are other (approximate) ways to calculate the \emph{integrated likelihood} and
the \emph{integrated classification likelihood}.
The Bayesian methods provide approximations that may, in practice, be at least
as good as the approximations that would be provided by methods such as the ICL.


The model-selection criterion $ILvb$ \cite{LatoucheILvbStatMod} is based on
a variational approximation to a fully Bayesian model.
As a result of its Bayesian model, it is an approximation of the integrated likelihood
and no further adjustment is required for model selection.
As with any variational Bayes method, we assume that the independence assumptions within the variational approximation
are a good approximation of the true posterior.
A second assumption made by those authors is that the Kullback--Leibler divergence, the difference between the true posterior
and the variational approximation, is independent of $K$.
If these two assumptions hold, then the measure they use, which they call the $ILvB$, is equivalent
to $\mathrm{P}(x | K)$, the \emph{integrated likelihood}.
To select the number of clusters, they use that value of $K$ which maximizes the $ILvB$.

\section{Estimation}
\label{SECestimation}

In this section, we describe our MCMC algorithm which samples,
given a network $x$, from the posterior $K,z|x$.
The moves are Metropolis-Hastings moves \citep{HastingsMetropolis}.
We define the moves and calculate the proposal probabilities and close the section with a discussion of the label-switching
phenomenon, where we use the method proposed in \cite{NobileAllocationSampler}
to summarize the clusterings found by the sampler.

Our algorithm is closely based on the
\emph{allocation sampler} \cite{NobileAllocationSampler}, which was originally presented in the context of a mixture-of-Gaussians model.
In fact, it can be applied to any model that can
be collapsed to the form $\mathrm{P}(x,z,K)$ where $x$
is some fixed (observed) data and the goal is to sample the
clustering and the number of clusters $(z,K)$.

In the Gibbs sampler used in \cite{Nowicki-01}, the parameters are not collapsed, and
sampling is from 

\[
z,\pi,\theta | x,K .
\]


In their experiment on the Hansell dataset, $K$ was fixed to 2.
As a result of this value for $K$, $\theta$ reduced to a single real number
specifying the relative expected size of the two clusters.
Expressions were presented for $p( \theta | z,\pi,x,K)$, $P( z | \theta,\pi,x,K )$ and  $ p(\pi | z,\theta,x,K)$
such that the various elements $z_i$ (or $\pi_{kl}$) are conditionally independent of each other,
given $(\pi,x,K)$ (or $(z,x,K)$),
allowing for a straightforward Gibbs sampler.

In contrast, we develop an algorithm that searches across the full sample space of all possible clusterings, $z$, for all $K$, drawing from the posterior,

\[
z, K |  x ,
\]

\noindent using \cref{EQfinal} as the desired stationary distribution of the Markov Chain.

We use four moves:
\begin{itemize}
	\item \emph{MK}: Metropolis move to increase or decrease $K$, adding or removing an empty cluster.
	\item \emph{GS}: Gibbs sampling on a randomly-selected node. Fixing all but one node in $z$, select a new cluster assignment for that node.
	\item \emph{M3}: Metropolis-Hastings on the labels in two clusters. This is the M3 move proposed in \cite{NobileAllocationSampler}. 		Two clusters are selected at random and the nodes are reassigned to the two clusters using a novel scheme fully described in that paper. $K$ is not affected by this move.
	\item \emph{AE}: The \emph{absorb-eject} move is a Metropolis-Hasting merge/split cluster move, as described in \cite{NobileAllocationSampler}. This move does affect $K$ along with $z$.
\end{itemize}

At each iteration, one of these four moves is selected at random and attempted.
All the moves are essentially Metropolis-Hastings moves; a move to modify $z$ and/or $K$ is generated randomly,
proposing a new state $(z',K')$,
and the ratio of the new density to the old density $\frac{\mathrm{P}(z',K|x)}{\mathrm{P}(z,K|x)}=\frac{\mathrm{P}(x,z',K)}{\mathrm{P}(x,z,K)}$ is calculated.
This is often quite easy to calculate quickly as, for certain moves,
only a small number of factors in \cref{EQfinal} are affected by the proposed move.
We must also calculate the probability of this particular move being proposed,
and of the reverse move being proposed.
The \emph{proposal probability ratio} is combined with the \emph{posterior mass ratio}
to give us the move \emph{acceptance probability},
\begin{equation}
\operatorname{min} \left( 1,
\frac
{ \mathrm{P}(x,z',K') }
{ \mathrm{P}(x,z,K) }
\times
\frac
{ \mathrm{P}_\text{prop}( (K',z') \rightarrow (K,z)) }
{ \mathrm{P}_\text{prop}( (K,z) \rightarrow (K',z')) }
\right) \, ,
\label{EQdetailedbalance}
\end{equation}
where $\mathrm{P}_\text{prop}( (K,z) \rightarrow (K',z'))$ is the probability that the algorithm,
given current state $(K,z)$, will propose a move to $(K',z')$.

In the remainder of this section, we discuss the four moves in detail, derive the proposal probabilities
and describe the computational complexity of the moves.



\subsection{MK}

The \emph{MK} move increases or decreases the number of clusters by adding or removing
an empty cluster. 
If \emph{MK} is selected, then the algorithm selects with 50\%
probability whether to attempt to add an empty cluster, or to delete one.
If it chooses to attempt a delete, then one cluster is selected at random;
if that cluster is not empty, then the attempt is abandoned.
If it chooses to attempt an insert, it selects a new cluster identifier randomly from
$\{1,\dots,K+1\}$ for the new cluster and  inserts a new empty cluster
with that identifier, renaming any existing clusters as necessary.

The proposal probabilities are
\[
\begin{split}
	\mathrm{P}_\text{prop}( (K,z)  \rightarrow (K+1,z')) & = \frac {0.5} {K+1}
\\
\mathrm{P}_\text{prop}( (K',z')  \rightarrow (K'-1,z)) & =
\left\{
\begin{array}{rl}
	\frac {0.5} {K'} & \text{if } K' > 1
	\\  	0 & \text{otherwise}
\end{array}
\right. \, .
\end{split}
\]

By adding an empty cluster, $K$ increases to $K'=K+1$ and the posterior mass change is:
\[
\begin{split}
	\frac{\mathrm{P}(x,z,K')}{\mathrm{P}(x,z,K)} & =
\frac{K!}{(K+1)!}
\frac
{\left(
  \frac
{\Gamma(\alpha (K+1)) \prod_{k=1}^{K+1} \Gamma(n_k + \alpha)}
{\Gamma(\alpha)^{K+1} \Gamma(N + \alpha (K+1))}
\right)}
{\left(
  \frac
{\Gamma(\alpha K) \prod_{k=1}^K \Gamma(n_k + \alpha)}
{\Gamma(\alpha)^K \Gamma(N + \alpha K)}
\right)}
\\ & =
  \frac
{
  \Gamma(\alpha (K+1))
  \Gamma(N + \alpha K)}
{
  (K+1)
  \Gamma(\alpha K)
  \Gamma(N + \alpha (K+1))
}\,.
\end{split}
\]

The computational complexity of this move is constant.

\subsection{GS}
The Gibbs update move, \emph{GS}, selects a node 
$i$ at random to be assigned to a new cluster.
All other nodes are kept fixed in their current cluster assignment
i.e. a single element of the vector $z$ is updated. 
Denote by $z' = z_{\{z_i \rightarrow k\}}$ the modified clustering resulting from a move of node $i$ to cluster $k$.
For each possible value of 
 $ z_i \in \{1,\dots,K\} $, $z_i$ is chosen with probability proportional to $\mathrm{P}(x,z_{\{z_i \rightarrow k\}},K)$. 
The proposal is then accepted.
Bear in mind that this move often simply reassigns
the node to the same cluster it was in before the
\emph{GS} move was attempted.

The calculations involved in \emph{GS} are quite complex as
many of the factors in \cref{EQfinal} are affected.
The sizes of the clusters are changed as the node
is considered for inclusion in each cluster, and the
number of edges and pairs of nodes are changed in many of the blocks.
The computational complexity is $\mathcal{O}(K^2)+\mathcal{O}(N)$ as every block
needs to be considered for each of the $K$ possible moves
and every node may be checked to see if it is connected
or not to the current node.

The $\mathcal{O}(N)$ term is just a theoretical worst case over all possible networks.
Our algorithm iterates over the neighbours of the current node
and this is sufficient to perform all the necessary calculations.
There is no need to iterate over the non-neighbours and therefore
the average complexity is equal to the average degree,
which will be much less than $N$ in real-world sparse networks.
For small $K$k, and assuming a given average degree, the complexity of the \emph{GS} move is independent of $N$.

\subsection{M3}
\emph{M3} is a more complex move and was introduced in \cite{NobileAllocationSampler}.
Two distinct clusters are selected at random, $j$ and $k$.
All the nodes in these two
clusters are removed from their current clusters and placed in a list
which is then randomly reordered -- call this ordered list $A = \{a_1, \dots, a_{n_j+n_k}\}$, of size equal to the total number of nodes in the two clusters.
The software creates a temporary cluster to store these nodes until they are reassigned to the original two clusters.
One node at a time is selected from
$A$ and is assigned to one of the two clusters according to some assignment probability.
As the nodes are assigned (or reassigned) the new cluster assignments are stored in a list $B_h = \{b_1, \dots, b_{h-1}\}$, where $b_i$ is the new cluster assignment of node $a_i$ and $B_h$ represents the assignments before the $h^{\rm th}$ node in A is processed.

Iterating through the list $A$, $a_h$ is assigned to either cluster
 $j$ or cluster $k$ with probability satisfying

 \[ p^{a_h \rightarrow j}_{B_h} + p^{a_h \rightarrow k}_{B_h} = 1 \, , \]

\noindent conditional on the nodes $B_h$ that have already been (re-)assigned.
Conceptually, any arbitrary assignment distribution can be chosen,
as long as the probabilities
for each choice are non-zero and sum to one.
Once all nodes in the list have been assigned to the two clusters,
the proposal probability is given by

\[ \mathrm{P}_\text{prop}( z \rightarrow z') = \prod_{h =1}^{n_j + n_k} p^{a_h\rightarrow b_h}_{B_h} \, . \]

We remark that while the order in which the nodes are reinserted
is random, it can be shown that this random ordering does not 
affect the acceptance probability.

In \cite{NobileAllocationSampler}, it is proposed to choose the ratio of the assignment probabilities 
as the ratio of the two posterior probabilities resulting from the assignments of the first $h$ nodes.
Specifically, denote by $z_{\{a_h\rightarrow l, B_h\}}$, the clustering that assigns the first $h-1$ nodes of A according to $B_h$ and assigns $a_h$ to cluster $l$.
Let $P(x', z_{\{a_h\rightarrow l, B_h\}}, K)$ be the posterior probability of this clustering on the network $x'$ \emph{where all unassigned nodes and edges involving these nodes are ignored}.
Then
\[
	\begin{split}
	\frac
	{p^{a_h \rightarrow j}_{B_h}}
	{p^{a_h \rightarrow k}_{B_h}}
	=
	\frac
	{ \mathrm{P}(x',z_{\{a_h\rightarrow j, B_h\}},K) }
	{ \mathrm{P}(x',z_{\{a_h\rightarrow j, B_h\}},K) } \, .
	\end{split}
\]
This heuristic should guide the selection towards `good' choices.
To calculate the proposal probability of the reverse proposal,
the list A is again traversed  to calculate
\[ \mathrm{P}_\text{prop}( z' \rightarrow z) = \prod_{h =1}^{n_j + n_k} p^{a_h\rightarrow z_{a_h}}_{B'_h} \, , \]

\noindent where $B'_h = \{z_{a_1},\dots, z_{a_{h-1}}\}$.


Our algorithm has been optimized for sparser networks. 
The complexity of \emph{M3} is made up of three terms.
First, it is possible that many or all nodes will be reassigned, causing a complexity of $\mathcal{O}(N)$
while updating the data structure that records the size of each cluster.
Second, we keep a record of the number of edges within each block;
the M3 move will consider each edge in the network at most once, as it moves the edge from one block to another,
leading to a complexity of $\mathcal{O}(M)$, where $M$ is the number of edges in the network.
Finally, once the data structures have been updated, a new posterior mass must be calculated by iterating
over each cluster and over each block, querying the summary data structures, to sum the new terms in \cref{EQfinal};
this has a complexity of $\mathcal{O}(K^2)$.

Together, this gives a complexity of $\mathcal{O}(N)+\mathcal{O}(M)+\mathcal{O}(K^2)$.
The first term may be ignored, since for most networks that are considered here and in the literature, $M > N$.
As long as the number of clusters is small, $K^2 \ll M$, the  $\mathcal{O}(M)$ term dominates.
While in the worst case $M=N^2$, in practice, for the sparse networks we consider, $M\ll N^2$.  

\subsection{AE}

In the \emph{absorb-eject} \emph{AE} move, a cluster is selected at random
and split into two clusters, or else the reverse move can merge two clusters.
This move therefore can both change the number of clusters $K$ and
change the clustering $z$.
The move will first choose, with 50\% probability, whether to attempt a merge or split.

In the case of the split move,  one of the $K$ clusters is selected at random.
Also, the cluster identifier of the proposed new cluster is selected at random
from $\{1,\dots, K+1\}$.
Finally, the nodes in the original cluster  are assigned between the
two clusters.  This is similar to the \emph{M3} move and a heuristic to guide the assignment, as in \emph{M3}, could be considered. 
Instead, as in \cite{NobileAllocationSampler}, we use a \emph{probability of ejection}, $p_E$, selected randomly from a $\text{Uniform}(0,1)$
distribution, such that each node is
assigned to the new cluster with probability, $p_E$. In such as move, the proposal probability is dependent on $p_E$. Rather than specify an ejection probability, we integrate over the choice of $p_E$ in much the same manner as collapsing. 

Given $(z,K)$ and a proposal to split into $(z', K'=K+1)$,
where a cluster of size $n_k$ is split into clusters of size $n_{j_1}$ and $n_{j_2}$, the resulting proposal probability for an eject move is

\[
\mathrm{P}_\text{prop}( (z,K) \rightarrow (z',K') )
= 
\frac
{ \Gamma ( n_{j_1} + 1 ) \Gamma ( n_{j_2} + 1 ) }
{ K (K+1) \Gamma ( n_k     + 2                        ) } \,.
\]

For a merge, the proposal probability is simply obtained as the probability of selecting the two clusters for merger from the $K'=K+1$ possible clusters.
One cluster is selected which will retain its current nodes and which will expand to contain the nodes in another, randomly selected, cluster,

\[
\mathrm{P}_\text{prop}( (z',K') \rightarrow (z,K) )
= \frac1K \frac1{K+1} \,.
\]

The complexity is similar to that of the M3 move.

\subsection{Applying the moves}

In all simulations, discussed in \cref{SECevaluation}, the algorithm is seeded by initializing $K=2$ and assigning the nodes randomly to one of the two initial clusters.
The first two moves, \emph{MK} and \emph{GS}, are sufficient to sample the space but have slow mixing.
The \emph{AE} move is sufficient on its own as it can add or remove clusters as well as move
the nodes to reach any $(z,K)$ state. In practice, we'll see in \cref{SECevaluation} that the
combination of \emph{AE} and \emph{M3} is good in the initialization stages to
burn-in to a good estimate of both $z$ and $K$ and lessen the dependence on the initialization.
It is possible to envisage many possible extensions to these moves. For example, a form of \emph{M3} could be made which selects three clusters to rearrange.
The \emph{AE} move could be extended to  include the assignment heuristic of the \emph{M3} move.

\subsection{Label Switching}
\label{SEClabelswitching}
For any given $z$, with $K$ clusters (assuming they are all non-empty), there are
$K!$ ways to relabel the clusters, resulting in $K!$ effectively equivalent clusterings.
The posterior has this symmetry and as the MCMC algorithm proceeds it often
swaps the labels on two clusters, in particular during
the \emph{M3} move.
This is known as the \emph{label switching phenomena}.
The posterior distribution for any $z_i | x,K$ assigns node $i$ to each of the $K$
clusters with probability $\frac1K$, so in the long run every node is assigned with equal probability to every cluster. While each $z_i$ is
uniformly distributed between 1 and K, the components of $z$ are dependent on each other and
pairs of nodes that tend to share a cluster will tend to have the same
values at their corresponding component of $z$.
Depending on the context, this may not be an issue of concern. For example, if the aim is to estimate $K$ or 
to estimate the probability of two nodes sharing the same cluster, see \cref{FIGKis2or3share},
or to estimate the size of the largest cluster,
then label switching is not a problem.



However, it sometimes is desirable to undo this label switching by relabelling the clusters, such that nodes are typically assigned to a single cluster identifier along with those other nodes that they typically share a cluster with. Such a relabelling can, for example, make it easier to identify the nodes which are not strongly
tied to any one cluster.  

We use the algorithm in \cite{NobileAllocationSampler} to undo the label switching by
 attempting to maximize the similarity
between pairs of clusterings, after the burn-in clusterings have been discarded. 
Given two $z$ vectors, at two different points in the Markov Chain, $t$ and $u$,
define the distance between them to be
\[
D(z^{(t)}, z^{(u)}) = \sum_{i=1}^N I(z^{(t)}_i,z^{(u)}_i) \, ,
\]
where $I$ is an indicator function that returns 0 if
node $i$ is assigned to the same cluster at point $t$ and point $u$;
and returns 1 otherwise.

For each $z^{(t)}$, consider $z^{(*t)}$, one of the $K!$ possible 
relabelled versions of $z^{(t)}$. 
The Markov Chain is run for $a$ iterations, discarding the
first $b$ iterations as burn-in.

Ideally, the goal is to find the relabelling that minimizes the sum
over all pairs of $u$ and $t$,
\[
\sum_{t=b}^a \sum_{u=t+1}^a D(z^{(*t)}, z^{(*u)}) \, ,
\]
but it is not computationally feasible to search across
the full space of all relabellings.
\margnote{More details, as requested by Rev\#2.}
Each state can be relabelled in approximately $K!$ different ways,
the precise number depends on the number of non-empty clusters.
There are $a$ states altogether, therefore the space of all possible relabellings
of all states will have $(K!)^a$ elements; this will be untractable for non-negligible $a$.
In our experiments, $a$ tends to be of the order of one million.

Instead, we use the \emph{online} algorithm proposed in \cite{NobileAllocationSampler}.
It first orders the states from the Markov chain by the number of non-empty clusters.  Then, it iterates through the states, comparing each state
to all the preceding relabelled states and relabelling the current state such that the
total distance to all the preceding relabelled states is minimized.

We will see in \cref{SECsurvey} how this algorithm helps to summarize
the output of the Markov Chain.
This algorithm is fast. On a 2.4~GHz Intel~Zeon in a server with 128GB RAM, it takes 43 seconds
to process the output of 1 million iterations of that data.  In comparison, it takes  610
seconds to run the SBM MCMC algorithm in order to get the
states to feed into the label-unswitching algorithm.
Note also that the algorithm doesn't take up much memory ---
even with a network with 10 million edges, the memory usage doesn't exceed 2GB.

Once the label-switched set of states is obtained,
a posterior distribution of the clustering
for each node, $ z_i | x $, can be calculated.
There is a similarity here with variational methods \cite{Daudin-08, LatoucheILvbStatMod}
as they model the posterior in this manner,
where each node's variational posterior is independent of the other nodes' variational posterior.
It may be interesting to compare these approximate posteriors to the approximate posterior
found by our method.

In
\margnote{\dots in particular, a discussion of loss.  We can't prove anything here.}
the experiments we perform later in \cref{SECevaluation,SECsurvey}, the vast majority of nodes are strongly assigned 
by this label-switching algorithm
to one of the clusters with at least 99\% probability in the posterior.
Therefore, the distance $D(.,.)$ between each state and this `summary' state is
usually quite small.
We take this as an indication that the online heuristic has done a reasonable job
of minimizing the distance between the states, at least for those networks.

\section{Evaluation}

\label{SECevaluation}

In this section we first look at experiments based on synthetic data and follow in the next section with an application of the collapsed SBM to a survey network gathered by one of the authors at a recent summer school.
The synthetic analysis proceeds by generating networks of various sizes from the model and examining whether the algorithm can correctly estimate the number of clusters and the cluster assignments.

As mentioned in the previous section, all our experiments are done on a
2.4~GHz Intel~Zeon in a server with 128GB RAM, and the memory usage never exceeded 2GB.


\subsection{Estimating z}
\label{SECestimatingz}

A 40-node directed, unweighted network is generated from the model, containing 4 clusters of 10 nodes each.  
The block densities $\pi_{kl}$ are generated by drawing from a $\text{Uniform}(0,1) \equiv \text{Beta}(1,1) $ for each of the $4 \times 4 = 16$ blocks.

\begin{figure}[h!]
	\centering
\includegraphics[width=.8\columnwidth]{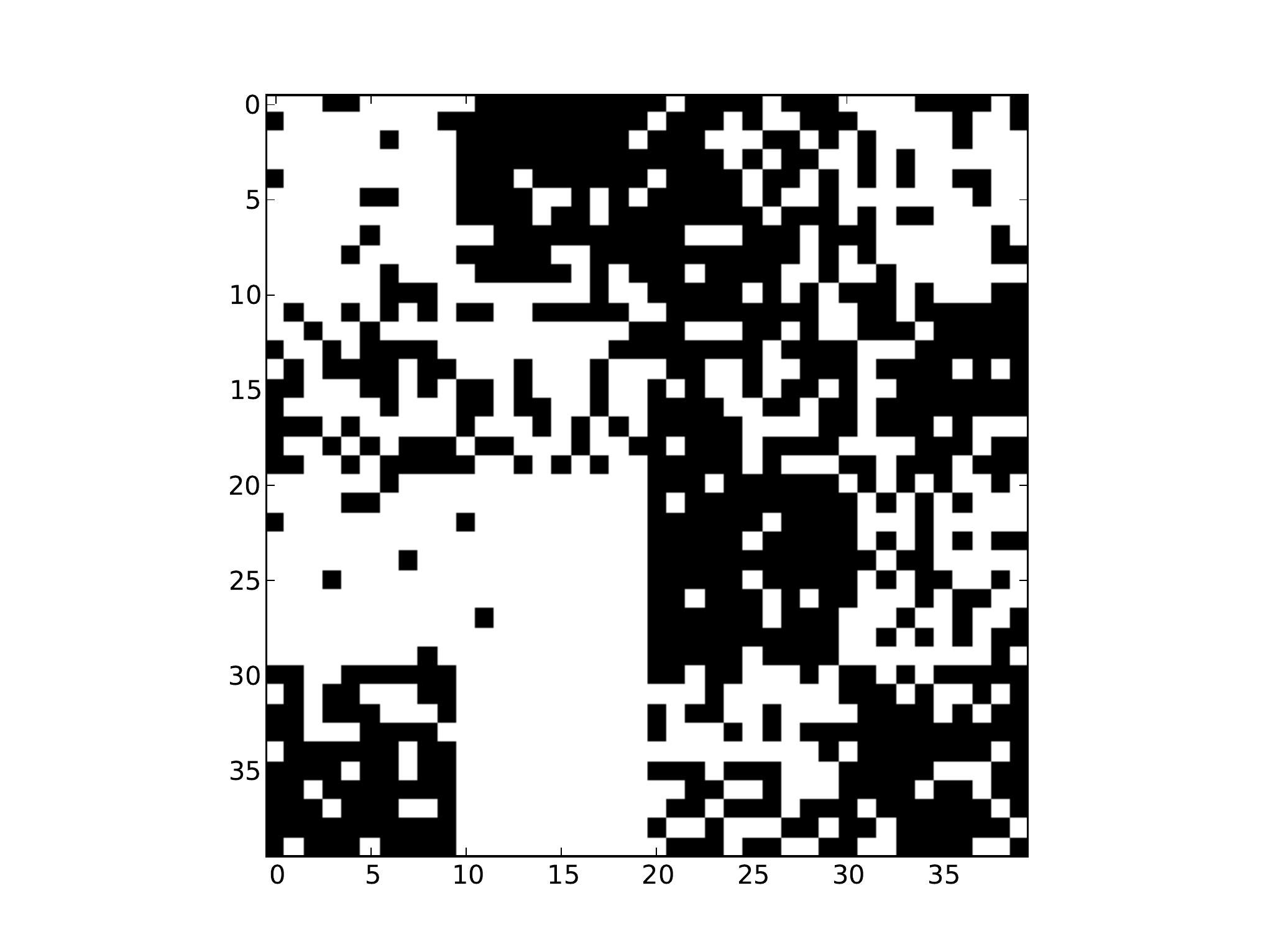}
\caption{The adjacency matrix (with $\delta = 0$) for the four-cluster synthetic network used in \cref{SECestimatingz}. Each of the four clusters has 10 nodes.}
\label{FIGK4_O50}
\end{figure}

To challenge the algorithm further we add noise to the synthetic data,
similar to simulation experiment described in section 4 of \cite{WyseFriel}.
The values in the matrix $\pi$ are scaled linearly. 
For a given $\delta$, define $\pi^{(\delta)}_{kl} = \delta+ \pi_{kl} (1-2 \delta)$.  While the values in the original $\pi$ are drawn from the full range, $[0,1]$, 
the elements in the matrix $\pi^{(\delta)}$ are in the range $(\delta, 1-\delta)$.
Various networks for values of $\delta$ between 0 and 0.5 are generated.
The original network model corresponds to $\delta=0$.
The network with $\delta=0.5$ corresponds to an Erdos-Renyi model with $p=0.5$ --- this is a random graph model with no block structure.

\begin{table}[h]
	\centering
	\begin{tabular}{r r r r r r}
		\hline
		$\delta$
		& \begin{turn}{80} $\mathrm{P}(K=4 | x)$ \; \end{turn}
			& $\hat{K}_\text{mode}$
			& \begin{turn}{80} $\mathrm{P}(K=\hat{K}_\text{mode} | x) $ \end{turn}
				& \begin{turn}{80} $\mathrm{P}(\hat{z}  \equiv  z | x)$ \end{turn}
		& $\tau$ \\
		\hline
		0.0  & 0.8982 & 4 & 0.8982 & 0.974 & 50.12 \\
		0.1  & 0.8799 & 4 & 0.8799 & 0.952 & 63.99 \\
		0.2  & 0.8769 & 4 & 0.8769 & 0.124 & 80.18 \\
		0.3  & 0.0073 & 2 & 0.7865 & 0.000 & 371.96 \\
		0.4  & 0.0075 & 1 & 0.6293 & 0.000 & 1365.58 \\
		\hline
	\end{tabular}
	\caption{The performance decreases as the noise level, $\delta$, increases. The fifth column, $\mathrm{P}(\hat{z} \equiv z|x)$, reports how often the sampler visits the `correct' answer;
	i.e. where the visited state was equivalent, subject to relabelling, to the model from which the network was generated. }
	\label{TBLdelta}
\end{table}

The algorithm is run for one million iterations, discarding the first 500,000 of these as burn-in.

\Cref{TBLdelta} shows how the performance is affected as $\delta$ increases.
The first column is the posterior probability for the ``correct'' answer for $K$, $\mathrm{P}(K=4|x)$.
As the value of $\delta$ increases, the network approaches the Erdos Renyi model and therefore
there is no longer any structure to detect; this explains why the accuracy decreases as
$\delta$ increases.
Next is the modal value of K which maximizes the posterior $\mathrm{P}(K|x)$, followed by the posterior probability of the modal value, $\mathrm{P}(K = \hat{K}_\text{mode}|x)$ .
The fifth column, $\mathrm{P}(\hat{z} \equiv z | x)$ is the probability that the (non-empty)
clusters are equivalent (allowing for relabelling) to the clustering used to generate the data.
Note that sometimes there are empty clusters in the estimate and therefore $\mathrm{P}(\hat{z} \equiv z | x)$ can be bigger than $\mathrm{P}(K=4 | x)$.

The final column reports $\tau$, the Integrated Autocorrelation Time (IAT) for the estimate of $K$, defined as $\tau = 1 + 2 \sum_{t=1}^{\infty} \rho(t)$, where $\rho_t$ is the autocorrelation at lag $t$. As the sampler visits the states, we consider how correlated the estimate of $K$ is with the estimates for preceding states.
A low autocorrelation, as summarized by the IAT, is an indicator of good mixing.

\subsection{Estimating K}
\label{SECestimatingK}

We perform three different types of experiments to judge the ability of the algorithm to correctly estimate the number of clusters
with networks of increasing size.

First, we repeat the experiments of \cite{LatoucheILvbStatMod}.
\margnote{These are the new requested experiments.  Our results are very similar to Latouche.}
The true numbers of clusters, $K_{true}$ is set to range from 3 to 7.  For each $K_{true}$, 100 networks
are randomly generated.  The number of nodes in each network, $N$, is set to 50.
The nodes are assigned to the clusters randomly, with $\theta_1= \dots= \theta_K = \frac{1}{K_{true}}$.
Two parameters are used to control the density of the blocks. The first,  
$\lambda$, is the density within clusters i.e. $\pi_{kk} = \lambda$.
Also, one of the clusters is selected to be a special cluster of `hubs', well connected to the other nodes in the network, by setting $\pi_{1k} = \pi_{k1} = \lambda$.
The second parameter, $\epsilon$, represents the inter-block density of all the other blocks i.e. $\pi_{kl}= \epsilon$ for $k, l\ne 1$.
As in the experiments of \cite{LatoucheILvbStatMod}, the parameter values
are $\lambda=0.9$, and $\epsilon=0.1$.

\begin{table}[h]
\centering
\subtable[ILvb \label{TBLilvb} ]{
	\begin{tabular}{|r|r r r r r|}
		\hline
		& \textbf{3} & \textbf{4} & \textbf{5} & \textbf{6} & \textbf{7} \\
		\hline
		\textbf{3} & 100    &0      &0      &0      &0 \\
  \textbf{4} & 0      &99     &1      &0      &0 \\
  \textbf{5} & 0      &4      &96     &0      &0 \\
  \textbf{6} & 0      &0      &24     &76     &0 \\
  \textbf{7} & 0      &5      &29     &41     &25 \\
		\hline
	\end{tabular}
}
\subtable[our algorithm \label{TBLSBMonLatoucheP} ]{
	\begin{tabular}{|r|r r r r r|}
		\hline
		& \textbf{3} & \textbf{4} & \textbf{5} & \textbf{6} & \textbf{7} \\
		\hline
		\textbf{3}    &99      &1       &0       &0       &0 \\
		\textbf{4}    &0       &99      &1       &0       &0 \\
		\textbf{5}    &0       &4       &96      &0       &0 \\
		\textbf{6}    &0       &0       &25      &75      &0 \\
		\textbf{7}    &0       &5       &27      &46      &22 \\

		\hline
	\end{tabular}
}
	\caption{
		The rows represent $K_{true}$ and the columns are the estimates from the $ILvb$ of \cite{LatoucheILvbStatMod} and from our algorithm.
	}
	\label{TBLbothLatoucheExpers}
\end{table}

Each network is run through the variational method of \cite{LatoucheILvbStatMod}.
The estimated value of $K$ which maximizes the $ILvB$ measure is taken as the estimate of the number of clusters.
A contingency table, showing the true number of clusters against the estimate from $ILvB$,
is displayed in \cref{TBLbothLatoucheExpers}~\subref{TBLilvb}.
For low $K_{true}$ the algorithm is very accurate, and for larger values
there is a tendency to underestimate the number of clusters.
For example, when $K_{true}=7$, the estimate was $\hat{K}=6$ for
41 of the networks and $\hat{K}=7$ for only 25 of the 100 networks.

The results from our algorithm, shown in \cref{TBLbothLatoucheExpers}~\subref{TBLSBMonLatoucheP} are similar to those obtained using the $ILvB$.

\subsection{Synthetic SBM networks}
\label{SECsynthSBM}
\margnote{Putting new experiments here within Latouche's framework.  Might remove our experiment.}

The experiments of \cref{SECestimatingK} involve synthetic data generated according to
a model of \emph{community structure}, where edges tend to form primarily within clusters.
In order to explicitly test our algorithm in the more general setting of \emph{block structure},
we generated another set of networks
with data generated directly from the SBM.

Similarly to the previous experiment, for each of a range of values of $K_{true}$, 100 networks are generated. $K_{True}$ is now set to range from 10 to 20 and the number of nodes is set to $N=100$, in order that the size of each cluster not be very small.  Each element
of $\pi_{kl}$ is chosen randomly from Uniform(0,1) and for each of the 100 networks, a new $\pi$ is created randomly.
As these are undirected networks, only the upper triangular portion of $\pi$ is used when generating the network.
Again, we compared the estimates of $K$ found by the $ILvB$ to those found by our algorithm.

\begin{table}[h!]
	\centering
	\begin{tabular}{|r|r r r r r r r r r r r|}
		\hline
		& \textbf{10} & \textbf{11} & \textbf{12} & \textbf{13} & \textbf{14} & \textbf{15} & \textbf{16} & \textbf{17} & \textbf{18} & \textbf{19} & \textbf{20} \\
		\hline
		\textbf{10}& \underline{72}&  15&   0&   1&   0&   0&   0&   0&   0&   0&   0 \\
		\textbf{11}& 15&  \underline{75}&   5&   3&   1&   0&   0&   0&   0&   0&   0 \\
		\textbf{12}& 5&  20&  \underline{64}&   6&   5&   0&   0&   0&   0&   0&   0 \\
		\textbf{13}& 2&   3&  21&  \underline{66}&   8&   0&   0&   0&   0&   0&   0 \\
		\textbf{14}& 0&   0&   4&  21&  \underline{61}&  10&   4&   0&   0&   0&   0 \\
		\textbf{15}& 0&   0&   2&   8&  28&  \underline{51}&   9&   0&   2&   0&   0 \\
		\textbf{16}& 0&   0&   1&   4&  15&  32&  \underline{33}&  11&   4&   0&   0 \\
		\textbf{17}& 0&   0&   0&   2&   4&  11&  30&  \underline{45}&   8&   0&   0 \\
		\textbf{18}& 0&   0&   0&   1&   3&  12&  20&  30&  \underline{23}&  10&   1 \\
		\textbf{19}& 0&   0&   0&   0&   0&   1&  12&  24&  38&  \underline{13}&  10 \\
		\textbf{20}& 0&   0&   0&   0&   0&   1&   7&   6&  23&  29&  \underline{23} \\
		\hline
	\end{tabular}
	\caption{The true number of clusters (rows) against the number estimated by $ILvb$ (columns).
		The diagonal entries are underlined to aid readability, as these represent the correct answer.
		We see here a tendency to underestimate the number of clusters,
		especially for larger $K_{True}$.
	}
	\label{TBLK10to20ILvB}
\end{table}
\begin{table} [h!]
	\centering
	\begin{tabular}{|r|r r r r r r r r r r r|}
		\hline
		& \textbf{10} & \textbf{11} & \textbf{12} & \textbf{13} & \textbf{14} & \textbf{15} & \textbf{16} & \textbf{17} & \textbf{18} & \textbf{19} & \textbf{20} \\
		\hline
		\textbf{10} &  \underline{95}&   0&   0&   0&   0&   0&   0&   0&   0&   0&   0 \\
		\textbf{11} &   6&  \underline{93}&   1&   0&   0&   0&   0&   0&   0&   0&   0 \\
		\textbf{12} &   1&   8&  \underline{90}&   1&   0&   0&   0&   0&   0&   0&   0 \\
		\textbf{13} &   0&   2&  12&  \underline{86}&   0&   0&   0&   0&   0&   0&   0 \\
		\textbf{14} &   0&   0&   1&   9&  \underline{90}&   0&   0&   0&   0&   0&   0 \\
		\textbf{15} &   0&   0&   0&   1&  13&  \underline{84}&   2&   0&   0&   0&   0 \\
		\textbf{16} &   0&   0&   0&   0&   1&  22&  \underline{73}&   4&   0&   0&   0 \\
		\textbf{17} &   0&   0&   0&   0&   0&   2&  29&  \underline{65}&   4&   0&   0 \\
		\textbf{18} &   0&   0&   0&   0&   0&   1&   9&  28&  \underline{62}&   0&   0 \\
		\textbf{19} &   0&   0&   0&   0&   0&   1&   3&   7&  38&  \underline{51}&   0 \\
		\textbf{20} &   0&   0&   0&   0&   0&   0&   0&   3&  11&  28&  \underline{57} \\

		\hline
	\end{tabular}
	\caption{The true number of clusters (rows) against the number estimated by our collapsed MCMC algorithm (columns).
		The diagonal entries are underlined to aid readability, as these represent the correct answer.
		The accuracy is better here than in \cref{TBLK10to20ILvB}; we can see that the numbers
		on the diagonal are larger.
	}
	\label{TBLK10to20SBMp}
\end{table}

The results  are shown in \cref{TBLK10to20ILvB,TBLK10to20SBMp}.
Each row of data represents the 100 networks generated for a given $K_{True}$.
Each column represents the estimated $\hat{K}$.
Ideally, the algorithm would correctly estimate the number of clusters
in most cases, corresponding to large number on the diagonal.  We have underlined the diagonal entries for clarity. Note that the sum of the entries in each row  does not always sum exactly to 100,
since there are cases where the algorithms underestimate or overestimate
the number of clusters, beyond the shown range.
For the $ILvb$ algorithm, it is necessary to specify a range of $K$ to be tested;
we specified the range from 5 to 30.
Our MCMC algorithm requires no such hint.

For networks with a small number of clusters, both algorithms perform well,
with 72\% accuracy for $ILvb$ and 95\% accuracy for our algorithm.
As the true number of clusters increase, the performance decreases.
Our algorithm maintains at least 50\% accuracy in all cases,
whereas the accuracy for $ILvb$ falls to 23\%.
When they are incorrect, both algorithms have a tendency to underestimate
the number of clusters.


In \cref{SEClargernetworks}, a more thorough investigation of  the speed and scalability of our algorithm
with respect to larger networks is given but we close our comparison with $ILvb$
with some remarks on speed.
For the first set of small networks above, both methods are very fast; 
they complete within seconds. For example, the $ILvb$ can be calculated
for all values of $K$ from 10 to 20 in a total under five seconds.
We have not defined a convergence criterion for our MCMC algorithm, and therefore
we make no attempt to halt the sampling early in order to define a `runtime' for our algorithm.
But in the occasions where both methods get the correct result, our algorithm
typically reaches the correct result within nine seconds;
and the sampler remains at, or very close to, the correct clustering for the remainder of the run.

\begin{figure}
	\centering
\subfigure[All moves enabled]{
\includegraphics[width=.45\columnwidth]{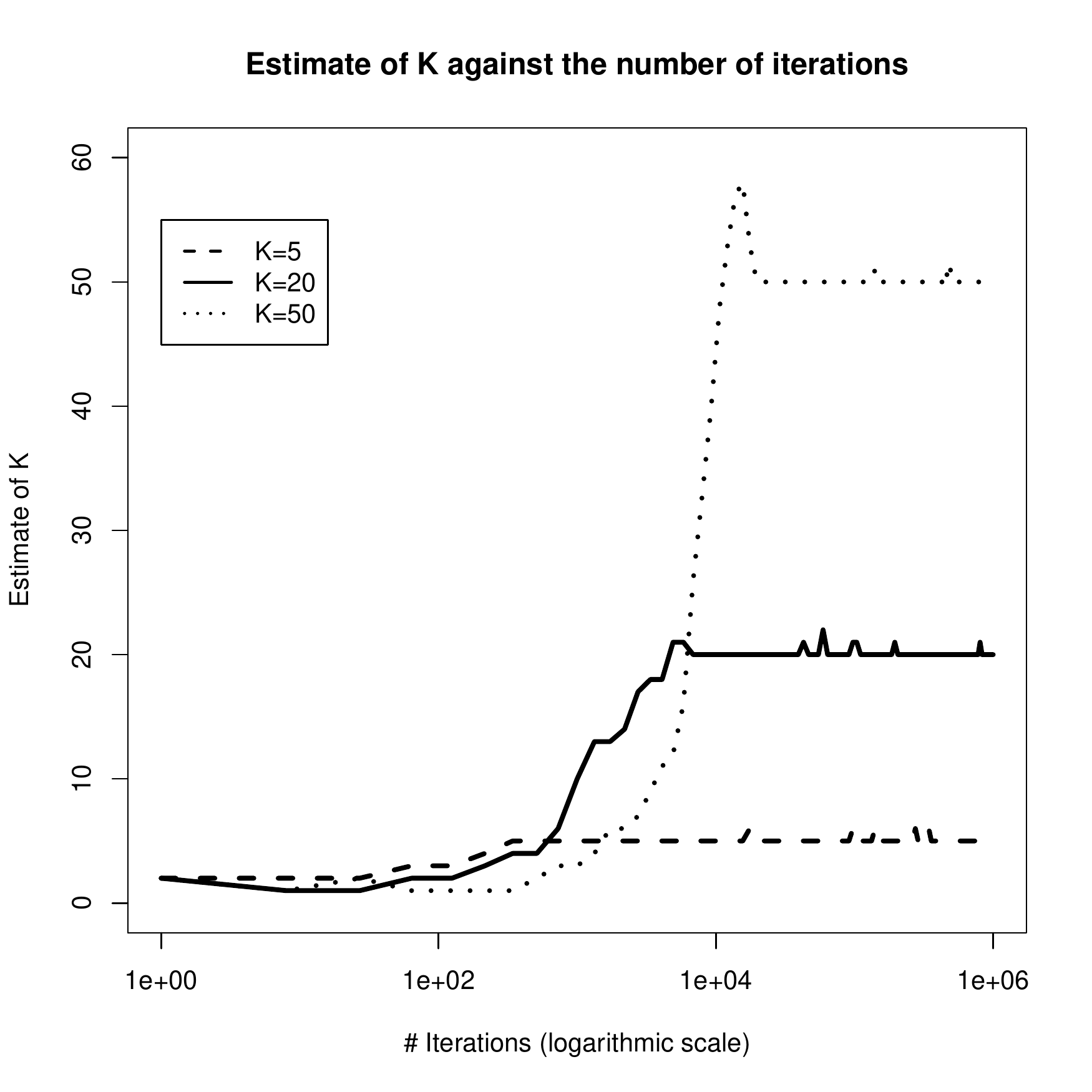}
}
\subfigure[\emph{AE} move disabled]{
\includegraphics[width=.45\columnwidth]{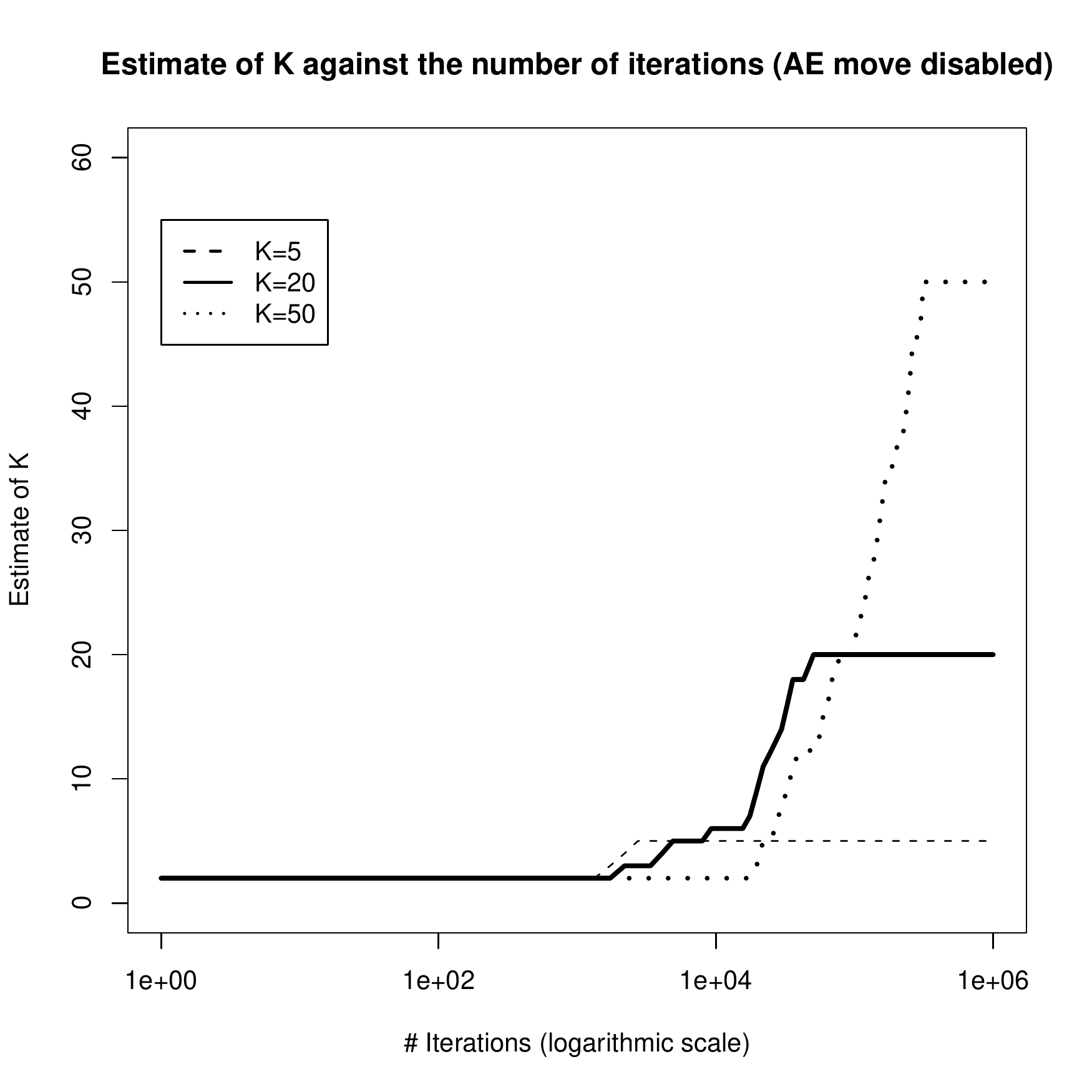}
\label{FIGsynthK_noAE}
}
\caption{The estimates of K in the synthetic networks, with $K = 5,20,50$.
The x-axis (logarithmic scale) is the number of iterations; as the algorithm proceeds, in each case
it converges on the correct estimate of $K$.
The networks had $10\times K$ nodes each.
In the lower plot, we see the performance where where the \emph{AE} move has been disabled; demonstrating how it is important in burnin.}
\label{FIGsynthK}
\end{figure}

Finally, to demonstrate the importance of the \emph{AE} move, in \cref{FIGsynthK}, the time taken by our algorithm to reach the correct clustering for three synthetic networks is shown.  The numbers of clusters in the networks is 5, 20, and 50 respectively, with $\pi$ drawn from $\text{Uniform}(0,1)$.  In each case, there were exactly 10 nodes in each cluster, giving $N=10 \times K$ nodes in each network.   
The x-axis displays the number of iterations and the y-axis the number of clusters at that stage
in the run of the sampler.
The correct clustering is reached in approximately 10,000 iterations.

We found that the \emph{AE} move is quite important, at least in the early stages.
If \emph{AE} is disabled, see \cref{FIGsynthK_noAE}, then it takes about 320,000 iterations
for K=50, instead of just 20,000 iterations when all moves are in effect.
For fast burn-in, \emph{M3} and \emph{AE} are necessary.
With similar experiments we noticed that, once the chain has burned in,
the \emph{M3} move is sufficient for good performance and the
other simple moves, \emph{GS} and \emph{MK}, do not make major contributions.

\subsection{Larger networks}
\label{SEClargernetworks}
Next, we investigate larger networks  to demonstrate
the scalability of the algorithm.

A number of synthetic networks are generated, each with approximately ten thousand nodes and ten million edges.
The number of clusters ranges from 3 to 50, and the number of nodes in each clusters, $O$, is set
such that the total number of nodes, $N = K \times O$, is close to 10,000.
If we use the default SBM edge model, then the number of edges would be approximately 50 million.
As this would take up a lot of computer memory, instead we modify the prior for the per-block densities to be Uniform(0,0.2) in order to ensure that the expected
number of edges is 10 million. Large real-world networks are typically quite sparse, even more sparse than this synthetic network.
The details, including the speed and accuracy, are in \cref{TBLlarger}.

The SBM algorithm is run for 100,000 iterations on each of these networks and the time to converge is recorded.
In each case, when the algorithm first visits the `correct' state, it remains in that state
for practically all the remaining iterations.
We record the number of iterations taken before the algorithm reaches the correct state,
and the time that has elapsed at that point.
It typically converges within one hour, but it takes nearly four hours for the 50-cluster network.
Methods that scale to thousands of nodes have been presented in the literature, such as
\cite{Daudin-08} and \cite{LatoucheILvbStatMod}. To our knowledge, ours is the only method
which has been demonstrated on networks with 10,000 or more nodes.

\margnote{More details here about speed versus ILvb}
We have attempted to load these networks into the $R$ software package in order to run them through $ILvb$.
However, the memory requirements for such large adjacency matrices become prohibitive.
For large networks, it may be necessary to consider a different implementation language and techniques
in order to fully explore the scalability of a variational method such as $ILvb$.
Instead, we generated five 500-node networks, with 20 clusters each, according to the SBM model and
run $ILvB$ on it, using only one value of $K$, namely $K=20$.
It takes between 38 and 56 seconds, depending on which of the five networks is used.
In comparison, on the same data, our algorithm takes between 17 and 35 seconds, despite the fact that it is given
no clue as to the correct value of $K$.
With 1,000-node networks, the runtimes for $ILvb$ are between 636 and 814 seconds,
whereas our algorithm takes between 55 and 78 seconds.
This suggests our algorithm scales better than the $ILvb$ -- although perhaps this
is an implementation issue rather than a limitation of the variational model.

In practice, it is necessary to run $ILvb$ for every possible value of $K$, and this fact
should be incorporated into any evaluation of its runtime.
For larger networks, the range of possible values of $K$ increases making this a
significant issue.
In contrast, an algorithm based on the allocation sampler, such as ours, does not suffer this limitation, suggesting that
that our algorithm is well suited to large networks.


\begin{table}
	\small
	\centering
	\begin{tabular}{ r r r r r r }
		\hline
		$K$ & $O$      & $N$    &          $E$ & $i$ & $t$
		\\ \hline
		3 & \numprint{3333}   & \numprint{9999} & \numprint{9722580}  & \numprint{41}     & \numprint{3317}
		\\ 4 & \numprint{2500}   &\numprint{10000} & \numprint{8526987}  &\numprint{149}     & \numprint{2977}
		\\ 5 & \numprint{2000}   &\numprint{10000} & \numprint{8627869}  &\numprint{190}     & \numprint{2460}
		\\ 6 & \numprint{1667}   &\numprint{10002} & \numprint{9974998}  &\numprint{416}     & \numprint{3265}
		\\ 7 & \numprint{1429}   &\numprint{10003} & \numprint{9316651}  &\numprint{749}     & \numprint{3449}
		\\ 8 & \numprint{1250}   &\numprint{10000} & \numprint{11059656} &\numprint{962}     & \numprint{3710}
		\\ 9 & \numprint{1111}   & \numprint{9999} & \numprint{9581440}  &\numprint{1383}    & \numprint{4052}
		\\ 10& \numprint{1000}   &\numprint{10000} & \numprint{9989886}  &\numprint{1277}    & \numprint{3785}
		\\ 20& \numprint{500}    &\numprint{10000} & \numprint{9871938}  &\numprint{5655}    & \numprint{4779}
		\\ 30& \numprint{333}    & \numprint{9990} & \numprint{9821594}  &\numprint{12497}   & \numprint{6999}
		\\ 40& \numprint{250}    &\numprint{10000} & \numprint{9862703}  &\numprint{37742}   & \numprint{12452}
		\\ 50& \numprint{200}    &\numprint{10000} & \numprint{10008963} &\numprint{40958}   & \numprint{24028}
		\\ \hline
	\end{tabular}
	\caption{
	The time-to-convergence for the larger synthetic networks.
	The networks have $N = K \times O$ nodes, made up of $K$ clusters each with $O$ nodes.
	After $i$ iterations ($t$ seconds), the algorithm reached the correct result and remained in, or close to, that state for the remainder of the 100,000 iterations. It should be noted that much of the runtime is simply taken up with loading the network into memory; the time spent in the MCMC algorithm itself is smaller than the $t$ figure presented here.
	}
	\label{TBLlarger}
\end{table}

\subsection{Autocorrelation in K}
\label{SECsmallAutoCorrs}

\tikzset{vertex/.style={black} }
\tikzstyle{abstract}=[rectangle, draw=black, rounded corners,
fill=white, drop shadow, text centered, anchor=north, text=black,
text width=3cm]
\tikzstyle{dot}=[fill=black,circle,minimum size=1pt]

\begin{figure}
	\centering
\subfigure[Adjacency matrix]{
	\tikz[scale=0.5] {
		\draw[help lines] (0,0) grid (6,6);
		\foreach \x / \y in {0/0,0/1,1/0,1/1,0/4,0/5,1/4,1/5,4/4,4/5,5/4,5/5}
		{
			\draw[fill=black] (\y,6-1-\x) rectangle +(1,1) {};
		}
	}
\label{FIGKis2or3}
}
\subfigure[Percentage posterior probability of two nodes sharing a cluster.]{
	\small
\begin{tabular}{|r|r|r|r|r|r|}
	\hline
	    & 97 &  4 &  4 & 75 & 75 \\ \hline
	 97 &    &  4 &  4 & 75 & 75 \\ \hline
	  4 &  4 & 99 & 99 &  4 &  4 \\ \hline
	  4 &  4 & 99 & 99 &  4 &  4 \\ \hline
	 75 & 75 &  4 &  4 &    & 97 \\ \hline
	 75 & 75 &  4 &  4 & 97 &    \\ \hline
\end{tabular}
\label{FIGKis2or3share}
}
\subfigure[Autocorrelation on $K$.]{
\includegraphics[width=.4\columnwidth]{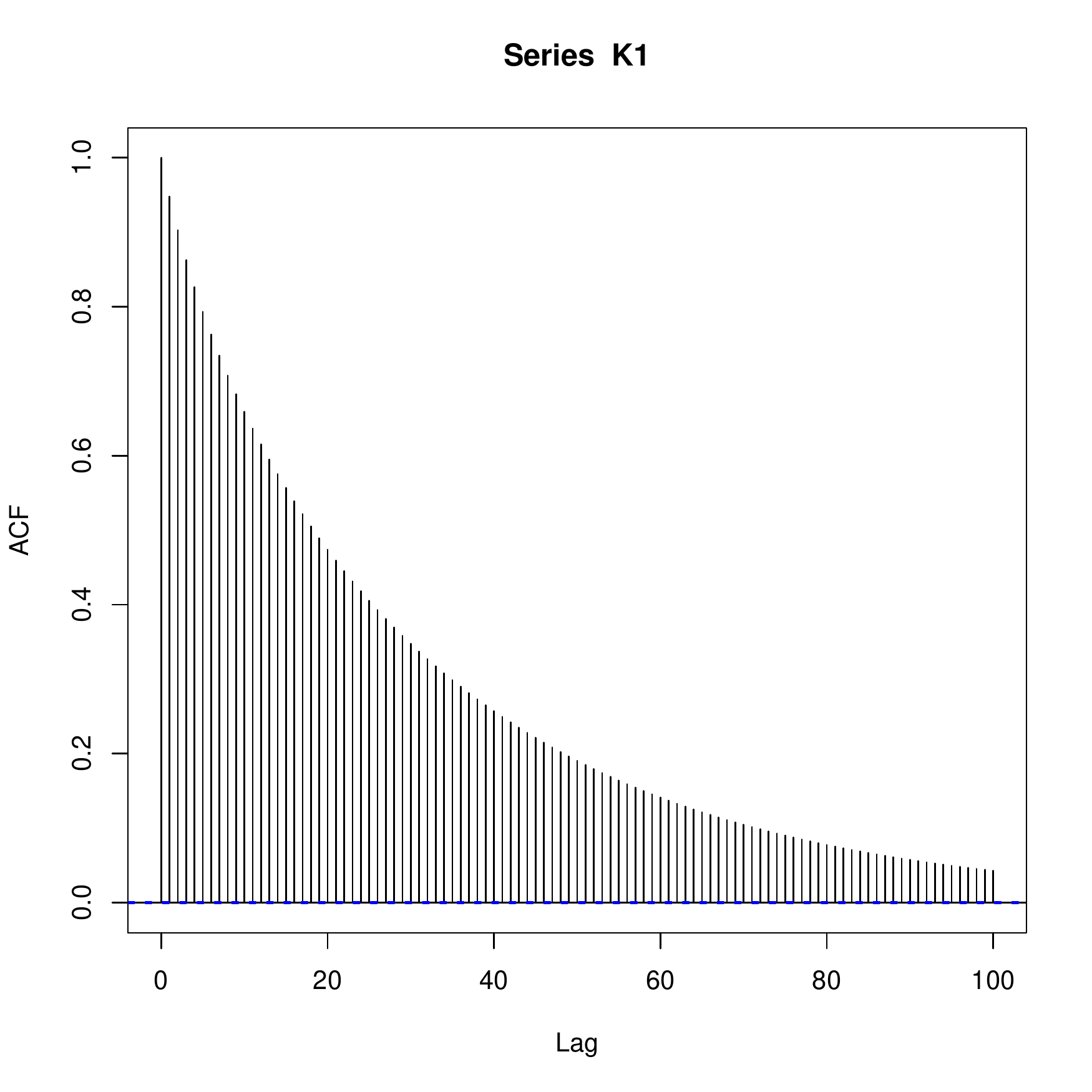}
\label{FIGKis2or3acor}
}

\caption{
  Adjacency matrix used in the analysis of varying K in \cref{SECsmallAutoCorrs}.
  \Cref{FIGKis2or3share} estimates, for every pair of nodes, the predicted probability of them sharing a cluster.
  \Cref{FIGKis2or3acor} shows the autocorrelation in the estimate of $K$.
}
\end{figure}

An autocorrelation analysis can reveal the mixing properties of the algorithm. However, in the above examples, and in the survey data discussed in \cref{SECsurvey}, the estimates of $K$ are very much peaked around
a single value. Often the larger values of $K$ are associated with empty clusters and
the estimate of the number of non-empty clusters  is even more peaked.
This makes it difficult to use $K$ as an interesting variable on which to perform autocorrelation analysis.
To address this, we examine the 6-node network in \cref{FIGKis2or3}, for which a greater variance in the values of $K$ is observed. Define $K_1$ to be the number of non-empty clusters, $K_1 \leq K$.
The posterior predictive probability for $K=2$ is 57.0\%, and for $K=3$ it is 31.4\%.
For the non-empty clusters, it is 73.4\% for $K_1=2$ and 24.4\% for $K_1=3$.
The autocorrelation in the estimates of $K$ is shown in \cref{FIGKis2or3acor}.

The acceptance rates on this small 6-node network are relatively high:
8.1\% for \emph{MK},
4.2\% for \emph{GS},
20.5\% for \emph{AE},
46.0\% for \emph{M3}
. We'll see lower
acceptance rates in the next section when  the algorithm is applied to the survey network.

\section{Survey of interaction data}
\label{SECsurvey}

A survey was performed by a team involving one of the authors of this paper at a summer school. 
We asked the 74 participants to fill in a survey and record which other participants they knew before the summer school and also
which participants they met during the school.  40 of the participants responded and gave us permission to make their survey response available publicly in anonymized format.
We created a directed, unweighted, network from the data by linking A to B if A recorded either type of relationship
with B, resulting in 1,138 edges. This network data is available at \url{https://github.com/aaronmcdaid/Summer-school-survey-network}.

\begin{centering}
\begin{figure}
\includegraphics[width=1.0\columnwidth]{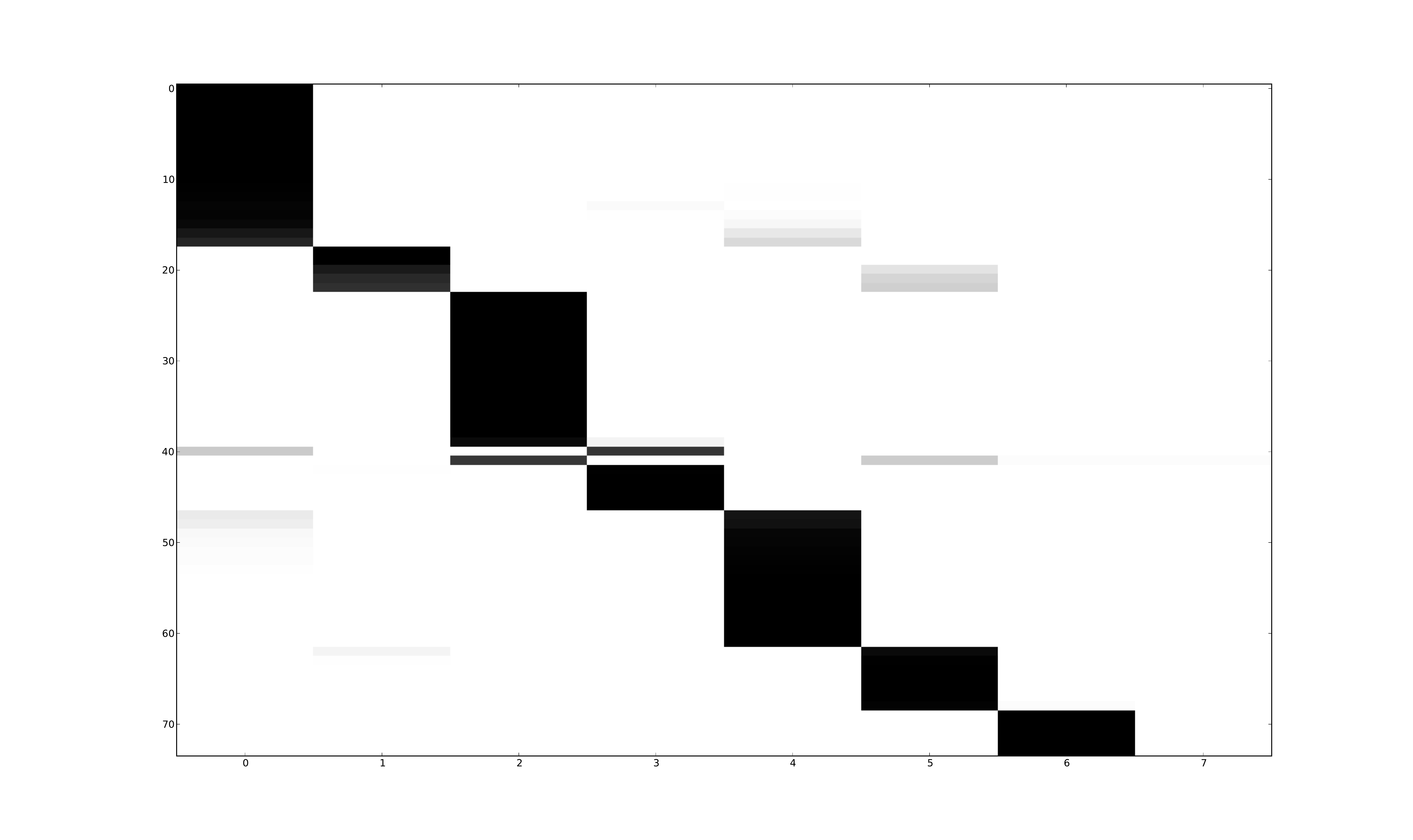}
\caption{The interation survey network of \cref{SECsurvey}. Node-to-cluster membership matrix.
74 rows, one for each participant. There are 8 columns, one for each of the
main seven clusters, and an extra cluster which, with very small probability,
is occupied by some nodes.
Most nodes are strongly assigned to one cluster, but the grey areas off
the diagonal show a small number of nodes that are partially assigned to
multiple clusters.
}
\label{FIGsurveyLabelSwitched}
\end{figure}
\end{centering}

Using the procedure described in \cref{SEClabelswitching}, we are able
to summarize the output of the Markov chain in \cref{FIGsurveyLabelSwitched}.
This is a matrix which records,
for each (relabelled) cluster and node, the posterior probability of
that participant being a member of that cluster.
Each row represents
one participant of the summer school, and the total weight in
each row sums to 1.0.
We have ordered the rows in this figure in order to bring similar rows together,
helping to highlight the sets of nodes which tend
to be clustered together in the Markov Chain.
As may be observed, most of the participants are strongly assigned
to one cluster. Every node is assigned to one of the clusters
with at least 75\% posterior probability, and the majority of nodes
have at least 99\% posterior probability.

\begin{figure}[ht!]
\includegraphics[width=1.0\columnwidth]{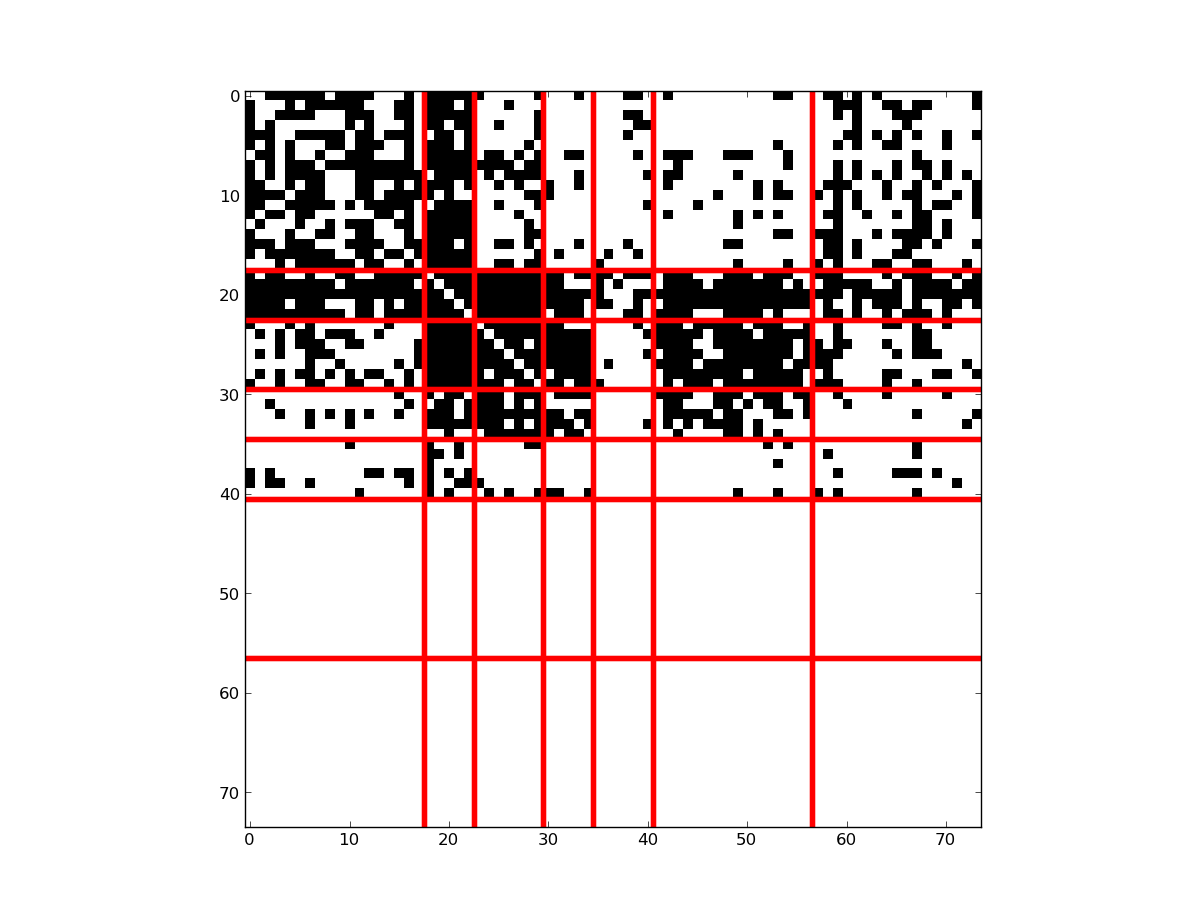}
\caption{The interaction survey network of \cref{SECsurvey} as a 74$\times$74 adjacency matrix for the 74 participants
in the summer school. 7 clusters were found by our method, and this matrix is ordered by the
summary clustering found by the label-unswitching method of \cref{SEClabelswitching}.
In the text in \cref{SECsurvey}, we interpret the clusters found and show how many of the clusters
correspond to the different types of people that attended the event.
There were 33 people who did not respond, these can be seen in the last two clusters.}
\label{FIGsurveyAdjOrdered}
\end{figure}

The number of clusters selected is 7, with 90.7 \% posterior probability.
We can summarize this into a single clustering by assigning each node
to its `best' cluster as found by the label-unswitching procedure.
In \cref{FIGsurveyAdjOrdered}, we see this clustering.
This particular clustering (or label-switched equivalents) has posterior probability of 20.7\%.
(The order in which the clusters are presented is different in \cref{FIGsurveyAdjOrdered} than 
in \cref{FIGsurveyLabelSwitched})


We then analyzed the clusters to see if they could be meaningfully interpreted.
The first thing that stands out is that the final two rows of blocks
are empty; these are simply the 33 people who did not respond to the survey.
It is interesting to see that the non-respondents have been split into two clusters.
Looking at the final two columns of blocks, the differences
in how other clusters linked to the non-respondents can be seen.

With the help of one of the organizers, we verified that the second cluster
(counting from the top, or from the left) is made
up of the \emph{Organizers} of the summer school, with one exception.
These were people based in the research institute who were involved in
organizing the summer school.
Therefore, it is no surprise that the corresponding rows and columns
of the adjacency matrix in \cref{FIGsurveyAdjOrdered} are quite dense.
The \emph{Organizers} interacted with almost everybody.

The third and fourth clusters are also made up of people who are based
in the research institute where the summer school was hosted
but who weren't on the programme committee.
We call these \emph{Locals}.
The first cluster is made up of \emph{Visitors}.
These were people from further afield who attended the school and spoke at the summer school.
Looking at the blocks at the top left of \cref{FIGsurveyAdjOrdered}
you can see that the \emph{Locals} know each other and the \emph{Visitors}
interacted with each other. But the two groups do not tend to interact strongly
with each other. The \emph{Organizers} are the glue that hold everybody together.
The fifth cluster appears simply to be made up of participants who did not
interact very much with anybody -- in fact they did not even interact with each other.

We can now interpret the fact that there are two clusters of non-respondents.
One of those clusters (the sixth cluster) is made of up of local
people. Their names appeared in the surveys of the \emph{Organizers} and \emph{Locals}.
The final cluster, the other non-respondent cluster, is made up of a broader range of people.
It includes many non-responder \emph{Visitors}, including many of the speakers at the summer school.

A community finding algorithm would not have been able to find these results, as
it would expect to find dense clusters and is tied to the
assumption that the probability of pairs of nodes being connected is, all other things being equal,
greater if they share a cluster than if they do not share a cluster.
This would manifest as dense blocks on the diagonal of this adjacency matrix.
Clearly, a community-finding algorithm could not find the non-respondent clusters.
Also, a community finding algorithm might have merged the \emph{Organizers} and \emph{Locals}
clusters. This is because those two clusters are quite dense internally and also have
many connections between them.
The only difference between these two clusters is how they interact with the rest of
the network;
this demonstrates how the rich block structure of the Stochastic Block Model,
including the various cluster-cluster interactions, can be helpful in clustering this data.

We ran the algorithm for 1 million iterations on this survey data,  discarding the first 500,000 iterations as burn-in.
The acceptance rates were as follows:
2.3\% for \emph{AE},
64.6\% for \emph{M3},
1.1\% for \emph{MK}.
In the case of the Gibbs sampler, 
2.5\% of the time it assigned a node to a new cluster,
otherwise the node was reassigned to its old cluster.

The \emph{M3} and \emph{AE} are both Metropolis-Hastings; a change to the
clustering is proposed and then the change is accepted or rejected.
Sometimes the accepted move actually places all the nodes back to
the same position they were in, or sometimes it merely swaps the labels
between the two clusters. If we consider these as `rejections',
then the rate and which new states are reached is just 1.0\% for \emph{M3}.
So, \emph{M3} is accepted a lot, but it usually only moves between
label-switched equivalents;
this tells us that the algorithm is able to move quickly between the various modes
of the distribution, and also suggests that the posterior is quite
peaked around the modes.

\subsection{Estimating the Network Probability, \texorpdfstring{$\mathrm{P}(x)$}{P(x)} }
\label{SECPx}
In Section 4, we discussed how the fully Bayesian approach of the SBM presented here allows model selection criteria such as the ICL to be avoided to select between models with different input numbers of clusters $K$.  It is also worth remarking that in certain circumstances, such as our survey data presented here, it is possible to compute an estimate of the network probability, $P(x)$; that is, the probability, given just the total number of nodes $N$, that the network $x$ is observed from an SBM.  This provides an absolute measure of the  fit of the SBM to the observed data and could be used to test the hypothesis that the data is drawn from an SBM against some alternative model.   

In the survey data there is one clustering where it, along with its
label-switched equivalents, take up 20.7\% of the posterior probability.  
Call this $\hat{z}$.  Thus we have a value
$\hat{z}$ which is visited very often by the sampler and this
allows an accurate estimate of $\mathrm{P}(K,\hat{z} | x)$ to be obtained using
\[
7! \times \mathrm{P}(K=7, z=\hat{z} | x) = 0.207\,.
\]

Now inserting $x$, $K$ and $\hat{z}$ into the expression for the joint distribution, an estimate of $P(x)$ can be obtained using
\[
	\mathrm{P}(x) \mathrm{P}(K=7,z=\hat{z}|x) = \mathrm{P}(x,z=\hat{z},K=7)\,.
\]

In the case of the survey data we obtain $\log_2\mathrm{P}(x) \approx -2,482$. To put some perspective on this value, we can compare with a model that selects $x$ uniformly at random from all possible directed networks over $N=74$ nodes. In this case, we obtain $\log_2\mathrm{P}(x) = -N(N-1) =-5,402$.
As a second alternative, if  $x$ were generated from an Erdos-Renyi model, averaged over all possible edge probabilities drawn uniformly at random, then 
$\log_2\mathrm{P}(x) \approx -4,130$.

\margnote{There was a section here about the hierarchical extensions - but I've deleted that for Rev\#2.}
\section{Conclusion}
\label{SECconclusion}

The original stochastic blockmodel was tested on a small network
with two clusters. We have shown how Bayesian models, collapsing,
and modern MCMC techniques can combine to create an algorithm
which can accurately estimate the clusters, and the number of clusters,
without compromising on speed.

It is sometimes stated
that MCMC is necessarily slower than other methods,
``effectively leading to severe size limitations (around 200 nodes)'' \citep{GazalVariational}.
The MCMC method we have presented scales to thousands of nodes,
and is more scalable than a recent variational method.
We do not claim that MCMC will always be necessarily faster than the alternatives, but we observe that comments on
the scalability of Metropolis-Hastings MCMC depends on the particular
model and on the particular proposal functions used.
It may be an open question as to which methods will prove to be most scalable in the long term,
as further improvements are made to all methods.

\margnote{I'll have to redo the conclusion in light of the many changes - ignore this for now.}
Our application to the survey data demonstrated that
\emph{block-modelling} can detect structure in networks
that might be missed by \emph{community-finding} algorithms.
Sometimes
the links between clusters are more interesting than the links
within clusters.

\subsection*{Acknowledgements}
This research was supported by Science Foundation Ireland (SFI) Grant No. 08/SRC/I1407 - Clique Research Cluster

\section*{Appendix A}
\label{SECappendixA}

Here, we describe the integrations which show that \cref{EQcollapsed2}
is equivalent to \cref{EQfinal}.

\subsection*{A.1 Collapsing $\theta$}
Here, we show how to calculate
\begin{equation}
\mathrm{P}(z|K) = \int_\Theta \mathrm{p}(z,\theta | K) \; \mathrm{d} \theta \, .
\label{EQappPz}
\end{equation}
This corresponds to the first integration expression in \cref{EQcollapsed}.
$z$ is a vector which records, for each of the $N$ nodes, which cluster it has been assigned to.
The probability for each cluster is in a vector $\theta$, where
\[1 = \sum_{k=1}^K \theta_k \, . \]
We integrate over the support of the Dirichlet distribution, which we have denoted with $\Theta$
in \cref{EQappPz},

\[ \theta \sim \mbox{Dirichlet}({\alpha,\alpha, \dots}) \, . \]

where we made the common simplification in our prior that all members of the vector $\alpha$ are identical; $\alpha_k = \alpha$.

$\theta$ is drawn from Dirichlet prior,
\[ \mathrm{p}(\theta) = \frac1{\mathrm{B}(\alpha)} \prod_{k=1}^K \theta_k^{\alpha_k-1} \, , \]
where the normalizing constant $\mathrm{B}(\alpha)$ is
\[ \mathrm{B}(\alpha) = \frac{ \prod_{k=1}^K \Gamma(\alpha_k) }{ \Gamma\left( \prod_{k=1}^K \alpha_k \right) } \, . \]
To collapse $\theta$, the expression for $\mathrm{P}(z | K)$ becomes the Multivariate P\'olya distribution. 
In the derivation, we have defined $n_k$ to be the number of nodes in cluster $k$, i.e.
\[n_k = \sum_{i=1}^N \left\{ \begin{array}{cc} 1 & \text{if} \; z_i = k \\ 0 & \text{if} \; z_i \neq k\end{array} \right. \, . \]

In the following expression, we will also find it useful to define another vector of length $K$,
\[ \alpha' = (\alpha_1 + n_1, \alpha_2 + n_2, \dots, \alpha_K + n_K) \, , \]

\[
\begin{split}
	\int_\Theta \mathrm{p}(z,\theta | K) \; \mathrm{d}\theta = & {} \int_\Theta \mathrm{p}(\theta | K)               \mathrm{P}(z | \theta , K) \; \mathrm{d}\theta
	\\ = & {} \int_\Theta \mathrm{p}(\theta | K) \prod_{k=1}^K \theta_k^{n_k} \; \mathrm{d}\theta
	\\ = & {} \int_\Theta \frac1{\mathrm{B}(\alpha)} \prod_{k=1}^K \theta_k^{\alpha_k-1} \prod_{k=1}^K \theta_k^{n_k} \; \mathrm{d}\theta
	\\ = & {} \int_\Theta \frac1{\mathrm{B}(\alpha)} \prod_{k=1}^K \theta_k^{\alpha_k+n_k-1} \; \mathrm{d}\theta
	\\ = & {} \frac{\mathrm{B}(\alpha')}{\mathrm{B}(\alpha)} \int_\Theta \frac1{\mathrm{B}(\alpha')} \prod_{k=1}^K \theta_k^{\alpha_k+n_k-1} \; \mathrm{d}\theta
	\\ = & {} \frac{\mathrm{B}(\alpha')}{\mathrm{B}(\alpha)} 
	\\ = & {} \frac{\Gamma(\sum_{k=1}^K \alpha_k)}{\Gamma(N+\sum_{k=1}^K \alpha_k)} \prod_{k=1}^K \frac{ \Gamma(n_k + \alpha_k) }{ \Gamma(\alpha_k) }
	\\ = & {} \frac{\Gamma(K \alpha)}{\Gamma(N+K \alpha)} \prod_{k=1}^K \frac{ \Gamma(n_k + \alpha) }{ \Gamma(\alpha) } \, .
\end{split}
\]

\subsection*{A.2 Collapsing $\pi$}

Now we look at the second integration expression in \cref{EQcollapsed}.
This describes how to calculate the probability of a network, $x$, given a clustering, $z$, and the number of clusters, $K$.

\[ \mathrm{P}(x|z,K) = \int_{\Pi} \mathrm{P}(x,\pi|z,K) \; \mathrm{d} \pi \, . \]

This depends on whether we're using the unweighted (Bernoulli) or integer-weighted(Poisson) model for edges.
It is also possible to allow real-valued weights with a Normal distribution and suitable priors,
an example of such a model is solved in Appendix B.2 of \cite{WyseFriel};
that paper is relevant for all the derivations here as the collapsing approach is quite similar
as in this paper.

The number of pairs of nodes in block between clusters $k$ and $l$ will be denoted $p_{kl}$ - for blocks on the diagonal $p_{kk}$
will depend on whether the edges are directed and on whether self loops are allowed; see \cref{EQpkkCountPairs} for details.
The relevant probabilities for a given block will be shown to be a function only of
$p_{kl}$ and of the total weight (or total number of edges) in that block. We'll denote this total weight as
\[
y_{kl} = \sum_{i,j | z_i = k, z_j=l} x_{ij} \, .
\]

In an undirected graph, we should consider each pair of nodes only once,
\[
y_{kl} = \sum_{i,j | i<j, z_i = k, z_j=l} x_{ij} \, .
\]

We are to calculate the integral for
a single block. $x_{(kl)}$ represents the submatrix of $x$ corresponding
to pairs of nodes in clusters $k$ and $l$.
Our goal is to simplify the expression such that there
there is one factor for each block,

\[
\begin{split}
	\mathrm{P}(x|z,K) & = \prod \mathrm{P}(x_{(kl)}|z,K)
	          \\      & = \prod \int \mathrm{P}(x_{(kl)},\pi_{kl}|z,K) \; \mathrm{d} \pi_{kl} \, .
\end{split}
\]

For directed graphs, the product is $\prod_{k,l}$, giving $K \times K$ blocks.
But for undirected graphs, the product is $\prod_{k,l|k \leq l}$, giving $\frac12 K(K+1)$ blocks.
The domain of the integration will be either $\int_0^1$ or $\int_0^\infty$, depending on which of the
two data models, unweighted or weighted, is in effect.

We'll first consider the unweighted (Bernoulli) model. The probability of a node in cluster $k$ connecting to a node in cluster $l$ is constrained by 
\[0 < \pi_{kl} <1 \, ,\]
and each element of $x_{(kl)}$ is drawn from a Bernoulli distribution with parameter $\pi_{kl}$,
\[
\mathrm{P}(x_{(kl)} | \pi_{kl} , z, K) = \pi_{kl}^{y_{kl}} (1-\pi_{kl})^{p_{kl}-y_{kl}} \, .
\]
The prior for $\pi_{kl}$ is a  Beta($\beta_1,\beta_2)$ distribution.
\[
\begin{split}
	\mathrm{P}(x_{(kl)}            |z,K) = & {} \int_0^1 \mathrm{p}(x_{(kl)} , \pi_{kl} |z,K) \; \mathrm{d} \pi_{kl}
	\\ = & {} \int_0^1 \mathrm{p}(\pi_{kl}) \; \mathrm{P}(x_{(kl)} | \pi_{kl} ,z,K) \; \mathrm{d} \pi_{kl}
	\\ = & {} \int_0^1 \frac{ \pi_{kl}^{\beta_1-1} (1-\pi_{kl})^{\beta_2-1} }{\text{B}(\beta_1,\beta_2)} \;  \pi_{kl}^{y_{kl}} (1-\pi_{kl})^{p_{kl}-y_{kl}} \; \mathrm{d} \pi_{kl}
		\\ = & {} \int_0^1  \frac{ \pi_{kl}^{y_{kl}+\beta_1-1} (1-\pi_{kl})^{p_{kl}-y_{kl}+\beta_2-1} }{\text{B}(\beta_1,\beta_2)} \; \mathrm{d} \pi_{kl}
		\\ = & {} \frac{ \text{B}(y_{kl}+\beta_1,p_{kl}-y_{kl}+\beta_2) }{\text{B}(\beta_1,\beta_2)} 
		\\   & {} \times \int_0^1  \frac{ \pi_{kl}^{y_{kl}+\beta_1-1} (1-\pi_{kl})^{p_{kl}-y_{kl}+\beta_2-1} }{  \text{B}(y_{kl}+\beta_1,p_{kl}-y_{kl}+\beta_2)  } \; \mathrm{d} \pi_{kl}
		\\ = & {} \frac{ \text{B}(y_{kl}+\beta_1,p_{kl}-y_{kl}+\beta_2) }{\text{B}(\beta_1,\beta_2)}  \, ,
\end{split}
\]
where $\mathrm{B}(\beta_1, \beta_2) = \frac{ \Gamma(\beta_1) \Gamma(\beta_2) }{ \Gamma(\beta_1+\beta_2) }$ is the Beta function.
This result is closely related to the Beta-binomial distribution.

Next, we'll consider the Poisson model for edges in more detail.
Again, we will see that $p_b$ and $y_b$ are sufficient for $\mathrm{P}(x_{(kl)} |  K, z)$.

In this integer-weighted model, an edge (or non-edges) between a
node in cluster $k$ and a node in cluster $l$ gets its weight from a Poisson distribution
\[
x_i|\pi_{kl} \sim \mbox{Poisson}(\pi_{kl}) \, ,
\]
and $\pi_{kl}> 0$.

This gives us, iterating over the pairs of nodes in the block,
\[
\mathrm{P}(x_{(kl)} | \pi_{kl} , z, K) = \prod_{i,j \in k,l} \frac{ \pi_{kl}^{x_{ij}} }{ x_{ij}! } \mbox{exp}(-\pi_{kl}) \, .
\]

We can combine this expression for every block,

\[
\begin{split}
	\mathrm{P}(x| \pi , z, K) & = \prod_{kl} \mathrm{P}(x_{(kl)} | \pi_{kl} , z, K)
	\\                        & = \prod_{kl} \prod_{i,j \in k,l} \frac{ \pi_{kl}^{x_{ij}} }{ x_{ij}! } \mbox{exp}(-\pi_{kl})
	\\                        & = \prod_{ij} \frac1{ x_{ij}! } \prod_{kl} \prod_{i,j \in k,l} { \pi_{kl}^{x_{ij}} } \mbox{exp}(-\pi_{kl}) \, .
\end{split}
\]

We can ignore the $\prod_{ij} \frac1{x_{ij}!}$, as one of those will be included for every pair of nodes in the network.
That will contribute a constant factor to \cref{EQfinal}; this factor will depend only on the network $x$, and
it will not depend on $K$ or $z$ or any other variable of interest, and hence we can ignore it for the purposes of \cref{EQfinal}.
Therefore, for our purposes we will be able to use the following approximation in the derivation
\[
\mathrm{P}(x_{(kl)} | \pi_{kl} , z, K) = \prod_{i,j \in k,l} { \pi_{kl}^{x_{ij}} } \mbox{exp}(-\pi_{kl}) \, .
\]

We'll place a Gamma prior on the rates,
\[ \pi_b \sim \mbox{Gamma}(s, \phi) \, . \]

\[
\begin{split}
	\mathrm{P}(x_{(kl)} & | z, K) = \int_0^\infty \mathrm{p} (x_{(kl)}, \pi_{kl} | z, K) \mathrm{d} \pi_{kl} \\ 
	& = {} \int_0^\infty \pi_{kl}^{s-1} \frac{ e^{- \pi_{kl} / \phi} } { \Gamma(s) \phi^s } \prod_{i,j \in k,l} \frac{ \pi_{kl}^{x_{ij}} }{ x_{ij}! } e^{-\pi_{kl}} \mathrm{d} \pi_{kl} \\ 
	= {} \prod & \frac1{x_{ij}!} \int_0^\infty \pi_{kl}^{s-1} \frac{ e^{- \pi_{kl} / \phi} } { \Gamma(s) \phi^s } \prod_{i,j \in k,l} { \pi_{kl}^{x_{ij}} } e^{-\pi_{kl}} \mathrm{d} \pi_{kl} \\ 
	& \propto {}                   \int_0^\infty \pi_{kl}^{s-1} \frac{ e^{- \pi_{kl} / \phi} } { \Gamma(s) \phi^s } \prod_{i,j \in k,l} { \pi_{kl}^{x_{ij}} } e^{-\pi_{kl}} \mathrm{d} \pi_{kl} \\ 
	& =       {} \int_0^\infty \pi_{kl}^{s-1 + \sum x_{ij}} \; \frac{ \exp(-\pi_{kl} p_{kl} - \frac{\pi_{kl}}{\phi}) } { \Gamma(s) \phi^s }     \mathrm{d} \pi_b  \, .
\end{split}
\]

We said earlier that we'll define $y_{kl} = \sum_{i,j \in k,l} x_{ij} $. We'll now substitute that in and also use the following definitions:
\[
\begin{split}
	s' & = s + y_{kl} \\
	\frac1{\phi'} & = p_{kl} + \frac1{\phi} \, .
\end{split}
\]

Where $\text{Gamma}(s,\phi)$ was the prior on $\pi_b$, $\text{Gamma}(s',\phi')$ is the posterior now that we have observed edges with total weight $y_{kl}$ between $p_{kl}$ pairs of nodes.
Returning to $f$, and rearranging such that we can cancel out the integral (because it is the integral of a Gamma distribution and hence it will equal 1),
\[
\begin{split}
	f(x_{(kl)} | z, K) & =    {}  \int_0^\infty \pi_{kl}^{s'-1} \; \frac{ \exp(- \frac{\pi_{kl}}{\phi'}) } { \Gamma(s) \phi^s }     \mathrm{d} \pi_{kl} \\ 
	& = {} \frac{ \Gamma(s') \phi'^{s'} }{  \Gamma(s) \phi^s  } \int_0^\infty \pi_{kl}^{s'-1} \; \frac{ \exp(- \frac{\pi_{kl}}{\phi'}) } { \Gamma(s') \phi'^{s'} }     \mathrm{d} \pi_{kl} \\ 
	& = {} \frac{ \Gamma(s') \phi'^{s'} }{  \Gamma(s) \phi^s  }            \\
	& = {} \frac{ \Gamma(s + y_{kl}) \left(\frac{1}{p_{kl} + \frac1{\phi} }\right)^{s + y_{kl}} }{  \Gamma(s) \phi^s  } \, .
\end{split}
\]

\bibliography{aaron}
\bibliographystyle{model2-names}



\end{document}